\documentclass[preprint,12pt,authoryear]{elsarticle}

\usepackage{amssymb}
\usepackage{amsmath}

\usepackage[utf8]{inputenc} 

\newcounter{thmc}
\newcounter{defc}

\newcounter{lemc}
\newtheorem{theorem}[thmc]{Theorem}

\newtheorem{lemma}[lemc]{Lemma}

\newtheorem{definition}[defc]{Definition}

\begin{document}

\begin{frontmatter}



\title{Quantum Uncertainty and Entropy} 


\author[1,2]{Giovanni Chesi} 
\author[1,2]{Lorenzo Maccone}

\affiliation[1]{organization={QUIT Group, Physics Department, University of Pavia},
            addressline={via Bassi 6}, 
            city={Pavia},
            postcode={27100}, 
            state={Italy},
}

\affiliation[2]{organization={INFN Sez. Pavia},
            addressline={via Bassi 6}, 
            city={Pavia},
            postcode={27100}, 
            state={Italy},
}

\begin{abstract}
We review the plethora of uncertainty relations that appear in quantum mechanics and their nuances. We present both foundational applications, e.g. in understanding and defining complementarity, and practical applications, e.g. in quantum metrology and cryptography. Both variance- and entropy-based uncertainties are covered here.
\end{abstract}




\end{frontmatter}



\section{Introduction}
\label{sec1}
Quantum uncertainty is the primary manifestation of complementarity: not all properties of a quantum system can be jointly sharply defined. The history of quantum uncertainty has been quite colorful. It started as a foundational principle for quantum mechanics, the uncertainty {\em principle}, but it was quickly reformulated into the, more general, superposition principle. So, it is now generally seen as a
consequence of the postulates of quantum mechanics, and can be derived from them: the uncertainty {\it relations}\footnote{Famously, in \citet{peres1993} the index entry for "uncertainty principle" provocatively points to itself and only uncertainty relations are considered there}. Also its interpretation
has seen a radical shift: interpreted initially by \citet{heisenberg1927} both in terms of {\em disturbance} and in terms of the impossibility of jointly defining sharply all possible values of properties of an object ({\em preparation}), now only the second interpretation is retained for the original uncertainty relations \citep{dariano2003}.

In this chapter we review the main results regarding uncertainty relations, both from a foundational perspective and from the point of
view of possible applications, e.g.~in the field of quantum technology. There are two main families of uncertainty relations: the
variance-based ones \citep{heisenberg1927,robertson1929,trifonov1994} and the entropic ones \citep{hirschmann1957,bialynicki1975,deutsch1983,maassen1988}.

The variance-based are expressed in terms of a minimum value for the product, or for the sum, of the variances of the observables considered. This minimum value typically depends on the state of the system. The existence of a nonzero minimum value implies that
incompatible observables cannot all be sharply determined: at least one (or more) of the variances are nonzero (at least when calculated
on the considered state, if the bound is state dependent). In the case of product uncertainties, the product of variances will be null if
only one of the variances is null, giving less information on the incompatibility of the observables than the sum-based ones. However,
the product ones are prevalent because the lower bound can be expressed in terms of easily calculated quantities: the expectation
values of commutators and anti-commutators of the involved observables. Their disadvantage is twofold. First of all they are
typically state dependent, so they do not really express a property of the involved observables. Second, they are dimensional quantities: the
variance of an observable depends on the observables (dimensionful) eigenvalues and not only on its eigenstates. However, the true origin
of quantum uncertainty, namely the Born rule probabilities, do not at all depend on the eigenvalues of the observables, but only on the
eigenvectors: $p(a)=|\langle\psi|a\rangle|^2$, where $|\psi\rangle$ is the state of the system, and $|a\rangle$ is the eigenstate of the observable relative to its $a$-th outcome.

This is the reason behind entropic uncertainty relations: they express the uncertainty in terms of the entropy of the distribution
$p(a)$. The entropy evaluates the spread of the distribution without any reference to the eigenvalues. Typically the known lower bounds to
the sum of entropies of observables are state independent, so they truly express a property of the involved observables.  The drawback of
such entropic uncertainties is that there is no simple general form that gives good state-independent lower bounds, and the derivation of lower bounds to the sum of entropies is involved and the bounds are often not tight.

Up to now we have considered uncertainty relations connected to two or more incompatible observables. However, in quantum mechanics not all measurable quantities are expressed in terms of observables: parameters such as time and interferometric phase do not have
associated observables in textbook quantum mechanics [observables {\em can} be associated to them if quantum mechanics is modified slightly \citep{page1983}]. Surprisingly, uncertainty relations for parameters exist, and they take the form of quantum Cramér-Rao bounds \citep{braunstein1996}. In the simple case of parameters whose translation is generated by a ``Hamiltonian'' $H$, these take the form of a lower bound to the product of the variance of $H$ and the root mean square error of the parameter.

The uncertainty relations, in their multiple versions, find today a multitude of applications that span from quantum cryptography \citep{bennett1984}, whose security proofs rests on various version of uncertainty relations \citep{tomamichel2015}, to quantum metrology \citep{giovannetti2004,giovannetti2006}, mostly based on quantum Cramér-Rao type bounds, to entanglement detection \citep{guhne2009}, where uncertainty relations are used as entanglement witnesses, and various others in quantum tech.

From a foundational point of view, arguably quantum uncertainty has lost their status of fundamental principle: the superposition principle having taken its place. The superposition principle asserts that the state of a quantum system can always be expressed as a (complex) linear combination of other valid states. In other words, states are vectors (rather, rays) in a linear vector space. As shown below, the uncertainty relations are a direct consequence of this principle, together with the Born rule that assigns probabilities to measurement outcomes.

In the rest of the paper, we explore why and how the entropy describes the uncertainty of a random process and the rich approach that it provides to the complementarity of quantum observables. In section~\ref{sec2} we revisit the variance-based approaches, starting from the very first uncertainty relations (sections~\ref{subsec21} and~\ref{subsec22}) to the most relevant extensions (sections~\ref{subsec23},~\ref{subsec24}, and~\ref{subsec25}). In section~\ref{sec3} we show the connection between entropy and uncertainty by introducing the most important notions of entropy in information theory, from the Shannon axiomatic derivation to the most modern generalizations. There, we also embed these notions of entropy into the quantum framework. Then, in section~\ref{sec4}, we address the entropic uncertainty relations. We start from the first formulations for the position and momentum observables and explore the extensions to generic dimensions and number of systems, and, in section~\ref{sec5}, to the presence of a memory. Finally, in section~\ref{sec6}, two applications where the entropic uncertainty relations have been crucial, i.e. cryptography and metrology.
%
%
%

\section{Variance-based uncertanties}

\label{sec2}
The postulates of quantum mechanics are grounded on a probabilistic description of the properties of physical systems. Even in the ideal case where experimental errors and noise are not considered, the outcomes obtained from measuring a quantum state need to be given within a statistical framework. A quantifier of the spread in the outcomes of the measurement of an observable $\hat{A}$ on a system in the state $|\psi\rangle$ is the \textit{variance} ${\rm var}(\hat{A})$. 
For an observable $\hat{A}$ with spectral decomposition  
\begin{equation}
\hat{A} = \sum_{a}a|a\rangle\langle a|,
\end{equation}
where $a$ and $|a\rangle$ identify the eigenvalues and the eigenvectors of $\hat{A}$, and a state $|\psi\rangle \in \mathcal{S}(\mathcal{H})$,
the variance is defined as
\begin{equation}
{\rm var}_{\psi}(\hat{A}) \equiv \langle \psi|(\hat{A} - \langle\hat{A}\rangle)^2|\psi \rangle = \langle \psi|\hat{A}^2|\psi\rangle - \langle \psi|\hat{A}|\psi\rangle^2,
\end{equation}
with 
\begin{equation}
\langle \hat{A} \rangle \equiv \langle \psi|\hat{A}|\psi\rangle = \sum_a a|\langle a|\psi\rangle|^2
\end{equation}
the mean value of $\hat{A}$.
In this sense, by definition, \textit{the variance quantifies the uncertainty of an observable as the average squared deviation from its mean value}. The \textit{standard deviation} $\text{sd}_{\psi}(\hat{A})$ is the square root of the variance.
\\
The variance-based URs that we inspect in sections~\ref{subsec21},~\ref{subsec23},~\ref{subsec24} and~\ref{subsec25} identfy the so-called \textit{preparation uncertainty}, where the quantum uncertainty stems from the complementarity of the observables involved in the UR. In other words, given two complementary observables $\hat{A}$ and $\hat{B}$, the preparation URs establish the impossibility of preparing a quantum state for which the uncertainties of $\hat{A}$ and $\hat{B}$ can both be arbitrarily small.
In section~\ref{subsec22} we briefly introduce a recent formulation of the URs addressing the uncertainty originated from disturbance and measurement and discuss their difference with respect to the preparation URs, starting from Heisenberg's conceptualization. \\
For further details, we remark that \citet{dodonov2025} has provided a review entirely dedicated to variance-based URs.

\subsection{Heisenberg, Robertson and Schr\"odinger formulations}
\label{subsec21}

Heisenberg's formulation is the very first attempt to formulate an uncertainty relation stemming from the existence of state superpositions. It was formulated for the observables position $\hat{Q}$ and momentum $\hat{P}$ as follows \citep{heisenberg1927}
\begin{equation} \label{heisenberg}
\text{var}(\hat{Q})\text{var}(\hat{P}) \geq \hbar^2/4.
\end{equation}
Heisenberg used an heuristic argument to prove the relation in Eq.~(\ref{heisenberg}), which identifies the conceptualization of the uncertainty principle, that we briefly report it in the next section. A rigorous proof was later given by \citet{kennard1927} and \citet{weyl1928}. 
This implication of quantum theory is a form of uncertainty with no classical analog: it suggests that some properties of specific pairs of quantum observables prevent the existence of a measurement that determines both jointly and sharply. Two years later, \citet{robertson1929} generalized it to arbitrary observables by proving the relation
\begin{equation} \label{robertson}
\text{var}_{\psi}(\hat{A})\text{var}_{\psi}(\hat{B}) \geq \frac{1}{4}\left|\langle \psi|[\hat{A},\hat{B}]|\psi\rangle\right|^2,
\end{equation}
where $\hat{A}$ and $\hat{B}$ are two observables, $[\hat{A}, \hat{B}] \equiv \hat{A}\hat{B} - \hat{B}\hat{A}$ is their commutator and $|\psi\rangle$ is the state of the system. Note that the commutator of two observables captures a fundamental quantum property of the observables, which is the overlapping of the eigenstates. Indeed, if $\hat{A}$ and $\hat{B}$ do not commute, then they do not share a set of common eigenvectors \citep{fano1971}.
\\

The argument by Robertson was revisited and simplified by \citet{schrodinger1930}, who eventually derived a more general relation. In particular, he applied the Cauchy-Schwartz inequality 
to the \textit{variance states}  $|\sigma_{\psi}(\hat{X})\rangle\equiv(\hat{X}-\mathbb{I}\langle\hat{X}\rangle)|\psi\rangle$ \citep{puri2001}, as follows
\begin{equation} \label{proofschrodingher}
\langle \sigma_{\psi}(\hat{A})|\sigma_{\psi}(\hat{A})\rangle\langle \sigma_{\psi}(\hat{B})|\sigma_{\psi}(\hat{B})\rangle \geq \left|\langle \sigma_{\psi}(\hat{A})|\sigma_{\psi}(\hat{B})\rangle\right|^2.
\end{equation}
The left-hand side reduces to the product of the variances of $\hat{A}$ and $\hat{B}$ on $|\psi\rangle$, namely $\text{var}_{\psi}(\hat{A})\,\text{var}_{\psi}(\hat{B})$. The right-hand side yields
\begin{equation} \label{lb}
\left|\langle \sigma_{\psi}(\hat{A})|\sigma_{\psi}(\hat{B})\rangle\right|^2 = \langle\hat{A}\hat{B}\rangle\langle\hat{B}\hat{A}\rangle - \langle\hat{A}\rangle\langle\hat{B}\rangle(\langle\hat{A}\hat{B}\rangle+\langle\hat{B}\hat{A}\rangle)+\langle\hat{A}\rangle^2\langle\hat{B}\rangle^2
\end{equation}
Schr\"{o}dinger also pointed out that the product $\hat{A}\hat{B}$ can be split into real and imaginary parts as follows
\begin{equation} \label{idea}
\hat{A}\hat{B} = \frac{1}{2}\left(\{\hat{A},\hat{B}\}+[\hat{A},\hat{B}]\right) \equiv \frac{\hat{C}_+ + \hat{C}_-}{2}
\end{equation}
where $\hat{C}_+\equiv \{\hat{A},\hat{B}\} = \hat{A}\hat{B} + \hat{B}\hat{A}$ is the anticommutator of $\hat{A}$ and $\hat{B}$. By using Eq.~(\ref{idea}), one finds that the lower bound in Eq.~(\ref{lb}) can be re-expressed as $\frac{1}{4}\left[\left(\langle \hat{C}_+\rangle - 2\langle\hat{A}\rangle\langle\hat{B}\rangle\right)^2 + \left|\langle \hat{C}_-\rangle\right|^2\right]$
thus obtaining the new UR
\begin{equation} \label{schrodinger}
\text{var}_{\psi}(\hat{A})\,\text{var}_{\psi}(\hat{B}) \geq \frac{1}{4}\left[\left(\langle \hat{C}_+\rangle - 2\langle\hat{A}\rangle\langle\hat{B}\rangle\right)^2 + \left|\langle \hat{C}_-\rangle\right|^2\right].
\end{equation}
Here we find a new term added to the squared mean commutator, which is the square of \textit{the arithmetic mean of the standard deviations} of $\hat{A}$ and $\hat{B}$, namely their \textit{covariance}, since
\begin{align}
\frac{1}{2}\left(\langle \hat{C}_+\rangle - 2\langle\hat{A}\rangle\langle\hat{B}\rangle \right) &= \frac{1}{2}\left(\langle(\hat{A}-\langle\hat{A}\rangle)(\hat{B}-\langle\hat{B}\rangle)\rangle + \langle(\hat{B}-\langle\hat{B}\rangle)(\hat{A}-\langle\hat{A}\rangle)\rangle\right) \nonumber \\
&\equiv \text{cov}_{\psi}(\hat{A}, \hat{B}) = \text{cov}_{\psi}(\hat{B}, \hat{A}).
\end{align}
Therefore, the new term is a measure of the correlations of the two observables. 
Schr\"{o}dinger's UR in Eq.~(\ref{schrodinger}) can be rewritten in a more compact form by using the \textit{covariance matrix} $\Sigma_{\psi}(\hat{A},\hat{B})$, defined as
\begin{equation} \label{covmat}
\Sigma_{\psi}(\hat{A},\hat{B}) \equiv \begin{pmatrix}
\text{var}_{\psi}(\hat{A})  & \text{cov}_{\psi}(\hat{A}, \hat{B})    \\
\text{cov}_{\psi}(\hat{A}, \hat{B}) & \text{var}_{\psi}(\hat{B})
\end{pmatrix}
\end{equation}
and reads
\begin{equation} \label{moderns}
\text{det}\left[\Sigma_{\psi}(\hat{A},\hat{B})\right] \geq \frac{1}{4}\left|\langle \psi|[\hat{A},\hat{B}]|\psi\rangle\right|^2.
\end{equation}
Interestingly, the difference between Robertson's and Schr\"{o}dinger's proof is the application, in the former, of the further inequality $|z|^2 = \left(\text{Re}[z]\right)^2 + \left(\text{Im}[z]\right)^2 \geq  \left(\text{Im}[z]\right)^2$, holding for all complex numbers $z$.
This explains why the correlation terms are missing in the Robertson formulation. 

\subsubsection{Minimum uncertainty states}

Next, we address the states saturating these URs, in the general form expressed in Eq.~(\ref{moderns}). As we have shown so far, this result is a direct consequence of the Cauchy-Schwartz inequality applied to the variance states $|\sigma_{\psi}(\hat{A})\rangle$ and $|\sigma_{\psi}(\hat{B})\rangle$, which is saturated if and only if $|\sigma_{\psi}(\hat{A})\rangle$ and $|\sigma_{\psi}(\hat{B})\rangle$ are linearly dependent, namely $|\sigma_{\psi}(\hat{A})\rangle = \alpha|\sigma_{\psi}(\hat{B})\rangle$, with $\alpha\in\mathbb{C}$. This condition is equivalent to requiring that the state $|\psi\rangle$ satisfies the eigenvalue equation
\begin{equation} \label{eqev}
\left(\hat{A} + \alpha\hat{B}\right)|\psi\rangle = \left(\langle \hat{A}\rangle +\alpha\langle \hat{B}\rangle \right)|\psi\rangle.
\end{equation}
These states are known as \textit{minimum uncertainty states}. For the uncertainty relation of $\hat{A}$ and $\hat{B}$, note that the minimum uncertainty states depend on the parameter $\alpha$ shaping the pertaining eigenvalue equation and, of course, on the observables, namely $|\psi\rangle \equiv |\psi(\alpha, \hat{A}, \hat{B})\rangle$. First, note that the operator $\hat{A} + \alpha\hat{B}$ is \textit{normal}\footnote{We recall that an operator $\hat{A}$ is normal if it commutes with its Hermitian adjoint, i.e. $[\hat{A},\hat{A}^{\dagger}] = 0$. Hermitian and unitary operators are special cases of normal operators.} only if $\alpha$ is a real number. Only a normal operator features orthonormal eigenstates, implying that the states $|\psi(\alpha)\rangle$ are not orthogonal if ${\rm Im}(\alpha) \neq 0$. Moreover, simple manipulations Eq.~(\ref{eqev}) yield the relations
\begin{align}
&{\rm var}(\hat{A}) = -\alpha\left(\langle \hat{A}\hat{B}\rangle - \langle \hat{A}\rangle\langle\hat{B}\rangle\right) = -\frac{\alpha}{2}\left(\langle \hat{C}_+\rangle + \langle \hat{C}_-\rangle -2\langle \hat{A}\rangle\langle\hat{B}\rangle\right)  \\
&{\rm var}(\hat{B}) = -\frac{1}{\alpha}\left(\langle \hat{B}\hat{A}\rangle - \langle \hat{A}\rangle\langle\hat{B}\rangle\right) = -\frac{1}{2\alpha}\left(\langle \hat{C}_+\rangle - \langle \hat{C}_-\rangle -2\langle \hat{A}\rangle\langle\hat{B}\rangle\right), 
\end{align}
which, by requiring that the variances are real quantities, identify two important properties of the minimum uncertainty states $|\psi(\alpha)\rangle$, namely
\begin{align}
&{\rm Re}(\alpha)\left|\langle \hat{C}_- \rangle\right| + {\rm Im}(\alpha)\left(\langle \hat{C}_+\rangle - 2\langle \hat{A}\rangle\langle\hat{B}\rangle \right) = 0 \label{condalfa} \\
&{\rm var}(\hat{A}) = |\alpha|^2\,{\rm var}(\hat{B}). \label{condvar}
\end{align}
From Eq.~(\ref{condalfa}), we see that the case where the minimum uncertainty states belong to an orthonormal set, i.e. ${\rm Im}(\alpha) = 0$, has $\left|\langle \hat{C}_- \rangle\right| = 0$, implying that for these states the UR in Eq.~(\ref{moderns}) is reduced to the condition of positive determinant of the covariance matrix and Robertson's relation in Eq.~(\ref{robertson}) is trivial. On the contrary,  if ${\rm Re}(\alpha) = 0$, then the observables $\hat{A}$ and $\hat{B}$ are uncorrelated, i.e. ${\rm cov}_{\psi}(\hat{A},\hat{B})=0$, and Schr\"{o}dinger's UR is reduced to Robertson's one. Finally, from Eq.~(\ref{condvar}), we find that the modulus of $\alpha$ identifies the ratio between the variances.
\\
A remarkable family of minimum uncertainty states for the observables $\hat{Q}$ and $\hat{P}$ \citep{schrodinger1926}, which is ubiquitous in quantum information and quantum optics, is the one of \textit{Gaussian states} \citep{schumaker1986,trifonov1994,puri2001,ferraro2005,fu2020}, namely the family of states whose Wigner function is Gaussian.
\\
Finally, we point out a major issue of the URs presented so far, expressed in the most general form by Eq.~(\ref{moderns}). We note that the lower bound identifying the quantum uncertainty depends on \textit{the mean value} of the commutator, namely it depends on the state $|\psi\rangle$ of the system yielding the bounded quantity $\Sigma_{\psi}$. Then, if the commutator is an operator, as for the generators of the $SU(n,m)$ algebras, one can find states annihilating the bound, such as the eigenstates of the commutator associated with null eigenvalues. In these cases, the URs would be converted into trivial inequalities, since the determinant of the covariance matrix is always positive. However, since the uncertainty is originated from the complementarity of the inspected observables, and one would like a relation that gives a bound that depends only on the observables and not on a specific system state,
 the UR in Eq.~(\ref{moderns}) can be improved, as we will see in the following.

\subsection{From Heisenberg's original formulation to Preparation and Measurement Uncertainty Relations}
\label{subsec22}
As mentioned in section~\ref{subsec21}, the seminal work originating scientific debate and research on the quantum uncertainty, as we know it today, is Ref.~\citep{heisenberg1927}. It is often overlooked that the primary goal in that paper did not deal with establishing a specific notion of uncertainty and that the relation in Eq.~(\ref{heisenberg}) came as a conjecture from a more general argument. Rather than that, Heisenberg's main idea was re-defining some basic notions of physics theories such that position, momentum, path and energy in light of the discontinuous nature that these quantities had shown in quantum mechanics. The \textit{gedankenexperiments} devised in that work aimed at deriving quantum uncertainty from the discontinuities in the energy transfer occurring in observation processes, i.e. experiments performed by an observer on a physical system.
\\
Among the most well-known of those \textit{gedankenexperiments}, we find the measurement of the position of an electron by means of a $\gamma$-ray microscope: due to the Compton effect, the determination of the position of the electron will necessarily alter its velocity, and, "\textit{by using the basic equations of the Compton effect}", one can retrieve the relation between the uncertainties of position and momentum in Eq.~(\ref{heisenberg}), which is introduced as
\begin{equation} \label{original}
\text{Precision}(\hat{Q})\cdot\text{Precision}(\hat{P}) \sim h,
\end{equation}
where the object "\textit{Precision}" does not identify any specific quantifier \citep{heisenberg1930}.
Crucially, Heisenberg reads this relation as an "\textit{intuitive clarification}" of the commutation equation $[\hat{Q},\hat{P}] = -i\hbar$. On the same line, he analyzed experiments in terms of other conjugated observables, always ending with the same expression of the uncertainties as in Eq.~(\ref{original}).
\\
This attempt of clarifying fundamental aspects of the new-born quantum physics has presented for the first time the problem of the uncertainty, but it has been problematic from the very beginning: in the paper itself, it is reported a criticism by Bohr on the construction and interpretation of the \textit{gedankenexperiments}. After decades of debates, we can say that Heisenberg's argument merges two fundamental facets of quantum uncertainty: the  \textit{complementarity} of quantum observables and the \textit{disturbance} of subsequent measurements.
\\
The connection with Bohr's complementarity \citep{bohr1928} is implied by the reference to the commutation relation of conjugated pairs. \citet{kennard1927} and \citet{robertson1929} started from this point to formalize Eq.~(\ref{original}) as a proper inequality, demonstrate it as an application of the Cauchy-Schwarz inequality, and generalize it to generic pairs of observables. But, since then, it is important to notice that this line of research has been focused on just one of the two aspects originally involved in Heisenberg's argument, i.e. the complementarity. This is the reason why these inequalities and all the generalizations that we will inspect in sections~\ref{subsec23},~\ref{subsec24},~\ref{subsec25} and~\ref{sec3} are known as \textit{preparation uncertainty relations}: they assess the impact the uncertainty in the preparation of an observable on the statistical dispersion of the measurement of the other \citep{peres1993}, but they say nothing about how the measurement of one "disturbs" the other. Then, the preparation URs address repeated measurements of an ensamble of equally prepared states, and the quantifiers of the uncertainties are uniquely determined by \textit{a priori} probabilities, identified by the complementary properties of the observables.
\\
The notion of disturbance, on the contrary, refers to repeated measurements on the same quantum system. It comes from the interpretation that Heisenberg yields for his \textit{gedankenexperiments}: the determination of a quantity disturbs the following determination of the other. This is the understanding of quantum uncertainty that has been crystalized in the form of \textit{uncertainty principle} by \citet{ruark1928} and has erroneously been extended to the preparation URs. \citet{dariano2003} has clarified this point by distinguishing between uncertainty relations (meant as preparation URs) and uncertainty principle. In order to take into account for the disturbance, new URs have been introduced throughout the last twenty years, known as \textit{measurement uncertainty relations}. \citet{ozawa2003} addressed Robertson's UR purely in terms of noise and disturbance (namely, the noise in the measurement of one observable disturbs the measurement of the other) and showed that, in these terms, it could be violated. This observation led him to fomulate a new trade-off between noise and disturbance, which is probably the first measurement UR. The interpretation of these uncertainties has, however, been contentious \citep{werner2004,bush2007}. The idea has been motivated by the emergence of new techniques that, by exploiting the laws of quantum mechanics, allowed to perform measurements that minimally affect the system, such as \textit{triple-state measurements} \citep{ozawa2004} and \textit{weak measurements} \citep{lund2010}. The notion of Robertson's inequality as a preparation UR has then been reconsidered and, starting from this point, new measurement URs have been devised \citep{werner2004,dammeier2015}, also in terms of entropies (see section~\ref{sec4}).
\\
Complementarity and disturbance have eventually been recognized as two distinct key aspects of \textit{quantum incompatibility}, as a general property of collections of input-output devices, describing both state preparations (input) and measurements (device) \citep{heinosaari2016}. In this framework, the incompatibility is defined as the existence of measurements that cannot be performed jointly.
Interestingly, \citet{erba2024} has recently proved that the incompatibility implies the disturbance, but the converse does not hold: a classical theory, where all measurements are compatible, featuring the irreversibility of measurement disturbance can be constructed.

\subsection{Extensions to more than two observables}
\label{subsec23}
One of the major limitations of the first URs, introduced in section~\ref{subsec21}, is their definition for pairs of observables. However, there are sets of complementary observables with more than two elements. It is then legitimate to ask if URs for such setss can be devised. \\
Robertson first addressed this question \citep{robertson1934} and managed to prove a generalization of his result in Eq.~(\ref{robertson}) to the case of a generic \textit{even} number of observables, namely
\begin{equation} \label{rob2}
\prod_{j=1}^{2m}{\rm var}_{\rho}(\hat{A_j}) \geq \frac{1}{4^m}\left|{\rm Pf}_m(\langle [\hat{A}_r, \hat{A_s}]\rangle)\right|^2
\end{equation}
where $\rho$ is the state of the system and ${\rm Pf}_m(x_{jk})$ is the Pfaffian\footnote{The Pfaffian can be defined recursively for a $2m\times2m$ skew-symmetric matrix $X$ as the polynomial
\begin{equation}
\text{Pf}_{m}(X) = \sum_{j=2}^{2m}(-1)^jx_{1j}\text{Pf}_{m-2}(X_{1j})
\end{equation}
where $X_{1j}$ is obtained by removing the first and $j$-th rows and columns from $X$, with the convention that the Pfaffian of a $0\times 0$ matrix equals one.
} of the elements $x_{jk}$ of a $2m\times 2m$ matrix.
We recall that, for a skew-symmetric matrix $X$, we have ${\rm Pf}(X)^2 = \text{det}(X)$. Importantly, here we do not need the state to be pure, namely $\rho$ can be a generic mixed state.
\\
After this generalization by Robertson, the issue has not been revisited until \citet{sudarshan1992} addressed the minimum uncertainty states within a group-theoretic approach, which paved the way for the extension of Schr\"odinger's UR, in Eq.~(\ref{moderns}), to more than two observables, thus including the contribution of correlations \citep{sudarshan1995}. \\
These results, obtained in a quantum optics framework, were later generalized by \citet{trifonov1997,trifonov1998}. First, he noted that the Robertson bound for multiple observables in Eq.~(\ref{rob2}) could be expressed in terms of a generalized covariance matrix, thus keeping the contribution of the covariances, as follows
\begin{equation}\label{triforob}
\text{det}\left[\Sigma_{\rho}(\mathbf{\hat{A}})\right] \geq \text{det}\left[C_{-}(\hat{\mathbf{A}})_{\rho}\right],
\end{equation}
where $\Sigma_{\rho}(\mathbf{\hat{A}})$ is the covariance matrix in Eq.~(\ref{covmat}) generalized to the $2m$ observables $\mathbf{\hat{A}} = \{\hat{A}_j\}_{j=1}^{2m}$, and $C_{-}$ is the antisymmetric matrix of the mean commutators, namely $(C_{-}(\hat{\mathbf{A}})_{\rho})_{jk} = -i/2\langle[\hat{A}_j,\hat{A}_k]\rangle_{\rho}$. 
\\
From Eq.~(\ref{triforob}), one retrives the bound in Eq.~(\ref{rob2}) by taking uncorrelated observables and by noting that the determinant of an antisymmetric matrix can always be expressed as the squared Pfaffian of its elements. By setting $m = 1$, it reduces to Schr\"{o}dinger's UR, in Eq.~(\ref{moderns}). The re-formalization of the URs in terms of the covariance matrix was needed and has been crucial in a plethora of contexts, from assessing the geometrical properties of physical systems [e.g. $\Sigma_{\psi}$, for a pure state $|\psi\rangle$, is a metric tensor in the manifold of generalized coherent states \citep{provost1980}] to improving the performance of quantum information protocols [e.g. by using squeezed states to reduce the variance of a specific observable \citep{walls1983}].
\\
Furthermore, the inspection of the geometrical structure of $\Sigma_{\rho}$ allowed to assess new sets of URs \citep{trifonov1997,trifonov1998}. Indeed, by recasting the covariance matrix into a diagonal form, it is possible to compute its determinant as the product of the diagonal elements. We list here the most relevant properties of the covariance matrix.
\begin{enumerate}
\item $\Sigma_{\rho}(\mathbf{\hat{A}})$ is symmetric, by construction.
\item $\Sigma_{\rho}(\mathbf{\hat{A}})\geq 0 \,\forall\,\rho$, implying $\text{det}[\Sigma_{\rho}(\mathbf{\hat{A}})] \geq 0$.
\item As for the $2\times 2$ covariance matrix in Eq.~(\ref{covmat}), the diagonal elements of $\Sigma_{\rho}(\mathbf{\hat{A}})$ are the variances $\text{var}_{\rho}(\hat{A}_j)$ and the nondiagonal elements the covariances $\text{cov}_{\rho}(\hat{A}_j,\hat{A}_k)$.
\item Given a linear transformation $\Lambda$, invertible, with real entries and $\text{det}(\Lambda)=1$, on the set of $2m$ observables $\{\hat{A}_j\}_{j=1}^{2m}$, such that
\begin{equation}
\hat{B}_j = \Lambda \hat{A}_j,
\end{equation}
the covariance matrix $\Tilde{\Sigma}_{\rho}(\mathbf{\hat{B}})$ of the observables $\{\hat{B}_j\}_{j=1}^{2m}$ is given by
\begin{equation}
\Tilde{\Sigma}_{\rho}(\mathbf{\hat{B}}) = \Lambda \Sigma_{\rho}(\mathbf{\hat{A}}) \Lambda^T.
\end{equation}
\item $\text{det}[\Tilde{\Sigma}_{\rho}(\mathbf{\hat{B}})] = \text{det}[\Sigma_{\rho}(\mathbf{\hat{A}})]$.
\item The invariance of a given operator $\Theta$ under the action of $\Lambda$, i.e. $\Lambda \Theta \Lambda^T = \Theta$, implies the invariance of the quantity $\text{Tr}[\Theta\Sigma_{\rho}(\mathbf{\hat{A}})]$, namely
\begin{equation}
\text{Tr}[\Sigma_{\rho}(\mathbf{\hat{A}})\Theta] = \text{Tr}[\Tilde{\Sigma}_{\rho}(\mathbf{\hat{B}})\Theta].
\end{equation}
\end{enumerate}
In particular, we have $\Theta = \mathbb{I}$ for orthogonal transformations $\Lambda = \Lambda^T$, and 
\begin{equation} \label{simplesso}
\Theta = \Omega \equiv \begin{pmatrix}
0  & \mathbb{I}_m   \\
-\mathbb{I}_m & 0
\end{pmatrix}
\end{equation}
for symplectic transformations $\Lambda\Omega \Lambda^T = \Omega$. These two cases are important for what concerns the diagonalization of the covariance matrix. Indeed, for any state $\rho$, it is always possible to find an orthogonal matrix such that $\Tilde{\Sigma}_{\rho}(\mathbf{\hat{B}})$ is diagonal for some $\mathbf{\hat{B}}$, due to property 1. If $\Sigma_{\rho}(\mathbf{\hat{A}}) > 0$, then the Williamson theorem \citep{williamson1936,ferraro2005} implies that there is a symplectic transformation that can diagonalize it. But then, from Eq.~(\ref{covmat}), the symplectic diagonalization of the covariance matrix is ensured if $\text{det}[C_-] > 0$. \citet{trifonov1997} found that this condition is satisfied in the following case.
\begin{lemma}[\cite{trifonov1997}]
Given an even number of observables, satisfying the commutation relations
\begin{align} \label{comm}
&[\hat{A}_j,\hat{A}_{k+m}]=\delta_{j,k}[\hat{A}_j,\hat{A}_{j+m}] \\
&[\hat{A}_j,\hat{A}_{k}] = [\hat{A}_{j+m},\hat{A}_{k+m}] = 0 
\end{align}
with 
\begin{equation}
-i[\hat{A}_j,\hat{A}_{j+m}] >0,
\end{equation} 
$\forall \, j,k\in[1,m]$, then $\rm{det}[C_-(\hat{\mathbf{A}})] > 0$, and the covariance matrix is positive definite.
\end{lemma}
A basic instance where these relations describe a physical system is the case of the quadratures of the electromagnetic field, $\hat{Q}_j \equiv \hat{A}_j$ and $\hat{P}_j \equiv \hat{A}_{j+m}$.
\\
Hence, one can establish new URs that are invariant under symplectic transformations. First, by using property 6, one can recast $\Sigma_{\rho}(\mathbf{\hat{A}})$ into a diagonal form $\Tilde{\Sigma}_{\rho}(\mathbf{\hat{B}})$, with
\begin{equation}
\text{Tr}[(i\Sigma_{\rho}(\mathbf{\hat{A}})\Omega)^{2}] = \text{Tr}[(i\Tilde{\Sigma}_{\rho}(\mathbf{\hat{B}})\Omega)^{2}] = 2\sum_{j=1}^m\text{var}(\hat{B}_j)\text{var}(\hat{B}_{j+m}),
\end{equation}
where $\Omega$ is defined in Eq.~(\ref{simplesso}).
Then, by using the Robertson UR in Eq.~(\ref{robertson}) for each term of the sum, one gets the \textit{trace-invariant URs} \citep{trifonov1997}
\begin{equation} \label{trifo1}
 \text{Tr}[(i\Tilde{\Sigma}_{\rho}(\mathbf{\hat{A}})\Omega)^{2}] \geq 2\sum_{j=1}^m\left|\frac{1}{2}\langle [\hat{B}_j,\hat{B}_{j+m}]\rangle\right|^2,
\end{equation}
which, in the case $m = 1$, reduces to the Schr\"{o}dinger UR in Eqs.~(\ref{schrodinger}) and~(\ref{moderns}).
\\
The rich structure of the generalized UR in Eq.~(\ref{triforob}) allowed to derive further families of URs. Let us consider the characteristic equations for $\Sigma_{\rho}(\mathbf{\hat{A}})$ and $C_-$, assuming now a generic dimension $d$:
\begin{align}
&\text{det}\left[\Sigma_{\rho}(\mathbf{\hat{A}}) - \lambda\mathbb{I}_d\right] = \sum_{j=0}^{d}c_j(\Sigma)(-\lambda)^{d-j} = 0 \label{car1} \\
&\text{det}\left[C_- - \mu\mathbb{I}_d\right] =  \sum_{j=0}^{d}c_j(C_{-})(-\mu)^{d-j} = 0 \label{car2}
\end{align}
where $c_j$ are the characteristic coefficients of the pertaining matrices. The $j$-th characteristic coefficient can be expressed as the sum of the $j$-th order principal minors of the related matrix, namely the determinants of the $j\times j$ submatrices with $j \leq d$. Clearly, for a $d \times d$ matrix $M$, we have that $c_d(M) = \text{det}(M)$. But then, we can rewrite the UR in Eq.~(\ref{triforob}) as an inequality for the characteristic coefficients of order $d$, i.e. $c_d(\Sigma)\geq c_d(C_-)$. \citet{trifonov1998} proved that a similar inequality holds for all the characteristic coefficients of $\Sigma_{\rho}(\mathbf{\hat{A}})$ and $C_-$ in Eqs.~(\ref{car1}) and~(\ref{car2}), thus establishing the \textit{characteristic URs}
\begin{equation} \label{trifo2}
c_j(\Sigma)\geq c_j(C_-).
\end{equation}
These URs can be understood as bounds on the maximum precision that can be obtained when a subset of complementary observables $\{\hat{A}_i\}_{i=1}^j$ is measured out of the set $\mathbf{\hat{A}}=\{\hat{A}_i\}_{i=1}^d$. As for the generalized UR in Eq.~(\ref{triforob}) and the trace-invariant URs in Eq.~(\ref{trifo1}), the characteristic URs are invariant under linear orthogonal transformations.
\\
Among the addressed extensions to the Schr\"{o}dinger UR, just the Robertson ones in Eqs.~(\ref{rob2}) and~(\ref{triforob}) preserve the structure of the product of variances as a measure of the uncertainty, but they provide a non-trivial bound only if the number of observables is even. The case of three canonical observables has been addressed in Ref.~\citep{kechrimparis2014}, where specifically the observables position $\hat{Q}$, momentum $\hat{P}$ and the rotated position $\hat{R} \equiv -\hat{Q}-\hat{P}$ has been considered. There, it is established the tight UR
\begin{equation}
\text{var}_{\psi}(\hat{Q})\text{var}_{\psi}(\hat{P})\text{var}_{\psi}(\hat{R}) \geq \left(\tau\frac{\hbar}{2}\right)^3
\end{equation}
where the \textit{triple constant} $\tau$ equals $2/\sqrt{3}$. Interestingly, the authors pointed out that the repeated application of the Schr\"{o}dinger UR in Eq.~(\ref{schrodinger}) to the pairs $(\hat{Q},\hat{P})$, $(\hat{Q},\hat{R})$ and $(\hat{P},\hat{R})$ yields a relation that is never tight, namely there is no state that can saturate its bound. This implies that the URs for triples cannot be derived by generalizing the UR for canonical pairs.  In Ref.~\citep{ma2017}, the triple constant $\tau$ has been found to bound also the variance product of the spin components $\hat{S}_x$, $\hat{S}_y$ and $\hat{S}_z$, for which experimental verification has been also provided. Recently, two generalizations to an aribitrary number of observables have been found \citep{he2022,bhunia2025}. In particular, in Ref.~\citep{he2022}, it is proved a relation between the variance of an operator $\hat{A}$ and the \textit{numerical radius} of the commutator of $\hat{A}$ and the projector on the state of the system. Given a bounded linear operator $\hat{A}$ on the Hilbert space $\mathcal{H}$, its numerical range $\mathcal{R}(\hat{A})$ is defined as 
\begin{equation}
\mathcal{R}(\hat{A}) \equiv \left\{\langle \psi|\hat{A}|\psi\rangle : |\psi\rangle \in \mathcal{H}, ||\psi|| = 1\right\}
\end{equation}
and, hence, the numerical radius $r(\hat{A})$ is the largest element of $\mathcal{R}(\hat{A})$, namely
\begin{equation}
r(\hat{A}) \equiv \sup\left[|\lambda| : \lambda \in \mathcal{R}(\hat{A})\right].
\end{equation}
The UR found in Ref.~\citep{he2022} reads
\begin{equation} \label{he}
\prod_{j=1}^m \text{var}_{\psi}(\hat{A}_j) \geq r^2\left(\prod_{j=1}^m[\hat{A}_j,|\psi\rangle\langle\psi|]\right).
\end{equation}
Since the radius $r\left(\prod_{j=1}^m[\hat{A}_j,|\psi\rangle\langle\psi|]\right)$ can be analytically retrieved, one can find from Eq.~(\ref{he}) an UR for an arbitrary number $m$ of observables. Interestingly, two distinct bounds are obtained if $m$ is even or odd. Later, the URs in Eq.~(\ref{he}) have been sharpened in Ref.~\citep{bhunia2025} by exploiting an improvement of the Cauchy-Schwarz inequality.

\subsection{Extensions to more than one state}
\label{subsec24}

So far, we have shown how the notion of quantum uncertainty outlined in section~(\ref{subsec21})  has been extended to the case of more than two observables, measured on a same \textit{single} state $\rho$. \citet{trifonov2000} provided a further generalization, where observables measured on a set of \textit{different} states $\boldsymbol\rho\equiv\{\rho_j\}_{j=1}^n$ satisfy new URs. The intuition behind this extension grounds in the notion of complementarity and, in particular, in the superposition principle: given that the URs stem from the existence of overlapping states, then the variances obtained by measuring a same observable over different states that are not eigenstates of that observable should be related by a proper UR.
\\
Firstly, we note that, in the case of a single state $\rho$, there exists an alternative simple proof of the UR for many obervables in Eq.~(\ref{triforob}), which exploits the positivity of the \textit{Robertson matrix} $R(\hat{\mathbf{A}})_{\rho}$, defined as
\begin{equation} \label{robmat}
R(\hat{\mathbf{A}})_{\rho} \equiv \Sigma(\hat{\mathbf{A}})_{\rho} + iC_{-}(\hat{\mathbf{A}})_{\rho}.
\end{equation}
Namely, the UR in Eq.~(\ref{triforob}) is implied by $R(\hat{\mathbf{A}})_{\rho} \geq 0$. 
\\
\citet{trifonov2000} extended this condition to hold for a set of different states.
Indeed, the Robertson matrix in Eq.~(\ref{robmat}) can be straightforwardly generalized for a set of pure states $\boldsymbol\rho = \boldsymbol\psi \equiv\{\psi_j\}_{j=1}^n $, by considering the variance states introduced in section~\ref{subsec21}, e.g. $|\sigma_j(\hat{A}_j)\rangle = (\hat{A}_j - \mathbb{I}\langle\hat{A}_j\rangle)|\psi_j\rangle$, and taking the pertaining Gram matrix $R(\hat{\mathbf{A}},\boldsymbol\psi)$, such that $[R(\hat{\mathbf{A}},\boldsymbol\psi)]_{jk} = \langle\sigma_j(\hat{A}_j)|\sigma_k(\hat{A}_k)\rangle$. Hence, a first extension of the UR in Eq.~(\ref{triforob}) to more than one state can be found by requiring the positivity of $R(\hat{\mathbf{A}},\boldsymbol\psi)$ \citep{trifonov2000}, e.g.
\begin{equation} \label{stex1}
R(\hat{\mathbf{A}},\boldsymbol\psi) \geq 0,
\end{equation}
which is a family of URs for $m$ observables and $n$ states.
\\
However, it can be shown, by diagonalizing $R(\hat{\mathbf{A}},\boldsymbol\psi)$, that the URs retrieved from Eq.~(\ref{stex1}) with $n > 1$ factorize over distinct states. In order to obtain URs that cannot be reduced to the single-state inequlities addressed in the previous sections, one needs to consider characteristic URs [introduced in Eq.~(\ref{trifo2})] derived from Eq.~(\ref{stex1}), which yield two futher families of URs \citep{trifonov2000}, namely
\begin{align}
&\sum_{k=1}^nc_j\left[\Sigma(\hat{\mathbf{A}},\psi_k)\right]\geq \sum_{k=1}^nc_j\left[C_{-}(\hat{\mathbf{A}},\psi_k)\right] \quad \forall \, j \in [1,d] \label{stex2} \\
&c_j\left[\sum_{k=1}^nR(\hat{\mathbf{A}},\psi_k)\right] \geq \sum_{k=1}^nc_j\left[R(\hat{\mathbf{A}},\psi_k)\right] \quad \forall \, j \in [1,d]. \label{stex3}
\end{align}
The URs obtained from the inequalities in Eqs.~(\ref{stex1}),~(\ref{stex2}) and~(\ref{stex3}) are known as \textit{state extended URs}, and are classified according to the number of states $n$ and observables $m$ involved as \textit{URs of type (n,m)}. We report here some of the simplest specific cases investigated by \citet{trifonov2000}.
\\
URs of type $(1, m)$, as expected, yield either the generalized UR in Eq.~(\ref{triforob}) or the characteristic URs in in Eq.~(\ref{trifo2}), according to which family between the one in Eq.~(\ref{stex1}) and the one in Eq.~(\ref{stex2}) is addressed.
\\
It is interesting to note that, from Eq.~(\ref{stex1}), one can find out an UR for a single observable, given that more than one state is considered. In particular, the UR of type (2,1), for two states $|\psi_1\rangle$ and $|\psi_2\rangle$, reads
\begin{equation} \label{stex21}
\text{var}_{\psi_1}(\hat{A})\,\text{var}_{\psi_2}(\hat{A}) \geq \left|\langle\psi_1|\left(\hat{A}-\langle\psi_1|\hat{A}|\psi_1\rangle\mathbb{I}\right)\left(\hat{A}-\langle\psi_2|\hat{A}|\psi_2\rangle\mathbb{I}\right)|\psi_2\rangle\right|^2,
\end{equation}
unveiling correlations between the variances of $\hat{A}$ evaluated in different states. It is saturated if and only if $|\psi_1\rangle = |\psi_2\rangle$, which can be exploited to define a distance between pure states \citep{trifonov1999}.
\\
Type $(2,2)$ presents a richer scenario, yielding, for two states $|\psi_1\rangle$ and $|\psi_2\rangle$ and two observables $\hat{A}$ and $\hat{B}$, the following URs:
\begin{align}
&\text{var}_{\psi_1}(\hat{A})\,\text{var}_{\psi_2}(\hat{B}) \geq \left|\langle\psi_1|\left(\hat{A}-\langle\psi_1|\hat{A}|\psi_1\rangle\mathbb{I}\right)\left(\hat{B}-\langle\psi_2|\hat{B}|\psi_2\rangle\mathbb{I}\right)|\psi_2\rangle\right|^2 \label{stex221}\\
&\frac{1}{2}\left[\text{var}_{\psi_1}(\hat{A})\,\text{var}_{\psi_2}(\hat{B}) + \text{var}_{\psi_2}(\hat{A})\,\text{var}_{\psi_1}(\hat{B})\right] - \text{cov}_{\psi_1}(\hat{A}, \hat{B})\,\text{var}_{\psi_2}(\hat{A}, \hat{B}) \nonumber \\
&\geq \frac{1}{4}\langle\psi_1|[\hat{A},\hat{B}]|\psi_1\rangle \langle\psi_2|[\hat{A},\hat{B}]|\psi_2\rangle^*. \label{stex222}
\end{align}
The first one, Eq.~(\ref{stex221}), stems from the family in Eq.~(\ref{stex1}), while the second one, Eq.~(\ref{stex222}), can be derived either from the family in Eq.~(\ref{stex2}) or from the one in Eq.~(\ref{stex3}). While the UR in Eq.~(\ref{stex221}), like the one in Eq.~(\ref{stex21}), bounds the uncertainty by means of the correlations of the observables over different states, the UR in Eq.~(\ref{stex222}) leverages on the complementarity of $\hat{A}$ and $\hat{B}$ and is, therefore, the proper \textit{state extended Schr\"odinger UR}, as it generalizes to two states Eq.~(\ref{schrodinger}).
\\
The state-extended URs in Eqs.~(\ref{stex21}),~(\ref{stex221}) and~(\ref{stex222}) have been recast in tomographic representation for applications in quantum optics and quantum information \citep{chernega2011,chernega2012,chernega2012b} and, in particular, Eq.~(\ref{stex222}) have been experimentally verified by using optical homodyne tomography \citep{bellini2012}.

\subsection{Sum uncertainties}
\label{subsec25}
We have pointed out in section~\ref{subsec21} that the URs outlined so far feature a state-dependent bound, which can be trivial for specific choices of the state of the system, thus failing to witness the complementary properties of the involved observables. There is a further issue with the URs based on the product of variances: for discrete, finite-dimensional operators $\hat{A}_j$, the uncertainty quantifier itself $\prod_j\text{var}_{\psi}(\hat{A}_j)$ can be null if $|\psi\rangle$ is an eigenstate for one of the $\hat{A}_j$. The search for URs that better captures the concept of complementarity has led to consider the \textit{sum of the variances} as an uncertainty quantifier.
\\
URs based on the sum of variances can be understood from the geometric point of view in terms of the Pythagoras' theorem \citep{deguise2018}: the uncertainties play the role of the edges of a hypercube, whose diagonal is the total uncertainty, bounded by a positive constant.
\\
The first UR based on the sum of variances has first appeared in Refs.~\citep{trifonov2000,trifonov2002}, where the motivation for re-shaping the uncertainty quantifier stemmed from the need of re-visiting the definition of coherent states in terms of URs. Indeed, coherent states can be defined as the minimum uncertainty states for the Robertson UR in Eq.~(\ref{robertson}) with equal variances \citep{klauder1985,puri2001}. However, the assumption of equal variances can be removed in order to address a larger class of minimum uncertainty states, given that, in the definition of coherent states, one replaces the Robertson UR with
\begin{equation} \label{trifosum}
\text{var}_{\psi}(\hat{A}) + \text{var}_{\psi}(\hat{B}) \geq \left|\langle[\hat{A},\hat{B}]\rangle\right|.
\end{equation}
This relation is a consequence of the Robertson UR and it is intrinsically less precise than that: the minimization of Eq.~(\ref{robertson}) implies the one of Eq.~(\ref{trifosum}), but the inverse is not true. Indeed, it can be derived by just observing that $[\text{sd}_{\psi}(\hat{A}) - \text{sd}_{\psi}(\hat{B})]^2 \geq 0$, and, hence,
\begin{equation}
\text{var}_{\psi}(\hat{A}) + \text{var}_{\psi}(\hat{B}) \geq 2\text{sd}_{\psi}(\hat{A})\text{sd}_{\psi}(\hat{B}) \geq \left|\langle[\hat{A},\hat{B}]\rangle\right|.
\end{equation}
The idea of bounding the sum of the variances has been rigorously addressed in Ref.~\citep{pati2007}, where the relation between the sum of the uncertainties and the uncertainty of the sum of observables has been established, and reads
\begin{equation} \label{pati}
\sum_{j=1}^m\text{sd}_{\psi}\left(\hat{A}_j\right) \geq \text{sd}_{\psi}\left(\sum_{j=1}^m\hat{A}_j\right).
\end{equation}
Therefore, the uncertainty on a linear combination of complementary observables is always smaller than the combination of the uncertainties of every single observable. This is indeed a geometric property of the quantum uncertainty formulated in terms of variance. By manipulating Eq.~(\ref{pati}), it is straightforward to find that, given a set of $m$ positive numbers $p_j$ such that $\sum_{j=1}^mp_j = 1$ and $0\leq p_j \leq 1 \,\forall\,j\in[1,m]$, then
\begin{equation}
\sum_{j=1}^mp_j\text{sd}_{\psi}\left(\hat{A}_j\right) \geq \text{sd}_{\psi}\left(\sum_{j=1}^mp_j\hat{A}_j\right),
\end{equation}
namely the \textit{uncertainty defined as standard deviation is a convex function}. By applying Eq.~(\ref{pati}) to two sets of observables $\hat{\mathbf{A}}$ and $\hat{\mathbf{B}}$ with the commutation relations in Eq.~(\ref{comm}) and fixing the commutators $[\hat{A}_j,\hat{B_j}] = i \hat{C}_- \, \forall\, j\in [1,m]$, one obtains the UR
\begin{equation}\label{patiur}
\left(\sum_{j=1}^m \text{sd}_{\psi}\hat{A}_j\right)\left(\sum_{j=1}^m \text{sd}_{\psi}\hat{B}_j\right) \geq \frac{m}{2}|\langle \hat{C}_-\rangle|,
\end{equation} 
which is stronger than Trifonov's UR in Eq.~(\ref{trifo1}) by a factor $m$, since the latter implies
\begin{align}
\left(\sum_{j=1}^m \text{sd}_{\psi}\hat{A}_j\right)^2\left(\sum_{j=1}^m \text{sd}_{\psi}\hat{B}_j\right)^2& \geq \left(\sum_{j=1}^m \text{var}_{\psi}\hat{A}_j\right)\left(\sum_{j=1}^m \text{var}_{\psi}\hat{B}_j\right) \nonumber \\
&\geq \sum_{j=1}^m \text{var}_{\psi}(\hat{A}_j)\text{var}_{\psi}(\hat{B}_j) \nonumber \\
&\geq \frac{m}{4}|\langle \hat{C}_-\rangle|^2. \nonumber
\end{align}
Specific URs for the angular momentum \citep{rivas2008} and spin \citep{ma2017} have been formulated, which, like the one in Eq.~(\ref{patiur}), feature a state-dependent bound. Among the first relations based on the sum of variances with a state-independent bound, a lower bound was found for an arbitrary number of generic observables in Ref.~\citep{huang2012}, which, however, requires to perform a maximization that cannot be solved analytically.  Interestingly, this bound is derived by making use of the entropic URs. We will come back to this in section~\ref{sec3}. 
\\
Stronger tight URs have been devised in Ref.~\citep{maccone2014}. They can be expressed in the compact form
\begin{equation} \label{maccone}
\text{var}_{\psi}(\hat{A}) + \text{var}_{\psi}(\hat{B}) \geq \max\left(\mathcal{B}_{\text I}, \mathcal{B}_{\text{II}}\right)
\end{equation}
with
\begin{align}
&\mathcal{B}_{\text I} =\left|\langle[\hat{A},\hat{B}]\rangle\right| + \left|\langle\psi|\hat{A}+i\hat{B} |\psi^{\perp}\rangle\right|^2 \label{b1} \\
&\mathcal{B}_{\text{II}} = \frac{1}{2} \text{var}_{\psi}(\hat{A}+\hat{B}) = \frac{1}{2}\left|\langle\psi^{\perp}_{\hat{A}+\hat{B}}|(\hat{A}+\hat{B})|\psi\rangle\right|^2, \label{b2}
\end{align}
where the states $|\psi^{\perp}\rangle$ and $|\psi^{\perp}_{\hat{A}+\hat{B}}\rangle$ are orthogonal to the state of the system $|\psi\rangle$, but, while the former can be arbitrarily chosen, the latter is constructed as a variance state with respect to the observable $\hat{A}+\hat{B}$, namely
\begin{equation}
|\psi^{\perp}_{\hat{X}}\rangle \propto (\hat{X}-\langle\hat{X}\rangle)|\psi\rangle.
\end{equation}
Note that both the bounds of the URs encompassed in Eq.~(\ref{maccone}) are strictly positive, even if the state is prepared in an eigenstate of $\hat{A}$ and $\hat{B}$. Ultimately, only a common eigenstate of $\hat{A}$ and $\hat{B}$ could cancel them. In particular, we recognize in the first term of  $\mathcal{B}_{\text{I}}$ the bound of the trivial UR in Eq.~(\ref{trifosum}), while $\mathcal{B}_{\text{II}}$ is a consequence of the convexity of the uncertainty in Eq.~(\ref{pati} and bounds the sum of the uncertainties with the uncertainty of the sum. Therefore, they both can detect the complementarity of observables also in the cases where the UR derived from the Robertson one (and related relations) cannot.
\\
The novel URs in Eq.~(\ref{maccone}) have been generalized to an arbitrary number of complementary observables in Ref.~\citep{chen2015}. Subsequently, a new UR has been derived in Ref.~\citep{chen2016} by generalizing the one with the bound $\mathcal{B}_{\text{II}}$ in Eq.~(\ref{b2}). It reads

\begin{equation}
\sum_{j=1}^m\text{var}_{\rho}(\hat{A}_j)\geq \left[\lambda_{\text{max}}(\tilde{R}(\hat{\mathbf{A}})_{\rho})\right]^{-1}\text{var}_{\rho}\left(\sum_{j=1}^m\hat{A}_j\right)
\end{equation}
where $\tilde{R}$ is the Robertson matrix in Eq.~(\ref{robmat}) with normalized elements, namely 
\begin{equation} \label{chen}
\tilde{R}_{jk} = \text{Tr}\left[\rho\frac{(\hat{A}_j - \langle\hat{A}_j\rangle\mathbb{I})(\hat{A}_k - \langle\hat{A}_k\rangle\mathbb{I})}{\sqrt{\langle\hat{A}_j^2\rangle - \langle\hat{A}_j\rangle^2}\sqrt{\langle\hat{A}_k^2\rangle - \langle\hat{A}_k\rangle^2}}\right] = \frac{\langle\hat{A}_j\hat{A}_k\rangle - \langle\hat{A}_j\rangle\langle\hat{A}_k\rangle}{\sqrt{\langle\hat{A}_j^2\rangle - \langle\hat{A}_j\rangle^2}\sqrt{\langle\hat{A}_k^2\rangle - \langle\hat{A}_k\rangle^2}},
\end{equation}
and $\lambda_{\text{max}}(\tilde{R})$ is the pertaining largest eigenvalue. Note that this UR holds for a generic state $\rho$, which does not need to be pure. It is shown to be tighter than the one given by Eq.~(\ref{b2}) since $\lambda_{\text{max}}(\tilde{R}) \leq 2$.

\section{Entropy as a measure of uncertainty}
\label{sec3} 

We now discuss the entropic uncertainties. Claude E. \citet{shannon1948} introduced the idea of \textit{entropy} as a measure of the information of a transmitted string $X$. In particular, if the elements of a string are taken from a given alphabet $\{n\}_{n=0}^N$ where the symbols are distributed according to a distribution $p(n)$, the entropy is the amount of information gained from each element of the string, after having received and read it. It can be measured in \textit{bits}, a unit measure introduced again by Shannon, as a a shortname for \textit{binary digit}. Every symbol of an alphabet appearing in a transmitted string can carry at most one bit.
\\
The choice of the name $entropy$ for this quantifier of information was prompted by von Neumann, as reported by Shannon himself \citep{mcirvin1971}:
\\
\\
"\textit{My greatest concern was what to call it. I thought of calling it 'information,' but the word was overly used, so I decided to call it 'uncertainty.' When I discussed it with John von Neumann, he had a better idea. Von Neumann told me, 'You should call it entropy, for two reasons. In the first place your uncertainty function has been used in statistical mechanics under that name, so it already has a name. In the second place, and more important, no one really knows what entropy really is, so in a debate you will always have the advantage.}"
\\
\\
Hence, it is clear the connection between this notion of entropy and uncertainty: the information gained after learning an element of a string can be also understood as the initial uncertainty about that element, and it is equivalent to the \textit{disorder} of an esemble expressed by the thermodynamic definition of entropy, namely the uncertainty on the number of microscopic configuations yielding the same macroscopic system.
\subsection{Notions of entropy}
\label{subsec31}
After Shannon's seminal works \citep{shannon1945,shannon1948,shannon1949}, many generalizations of this notion of entropy have been introduced and addressed, each of them capturing a specific nuance of the uncertainty on the output of a random process. All of them are constructed based on the requirements identified by Shannon for his measurement of the uncertainty. In the following, we start with the first fundamental results found by Shannon, which formalize a basic theory of information, and revise the most relevant generalizations with an impact on quantum uncertainty relations.

\subsubsection{Shannon entropy}
\label{subsubsec311}
In order to understand why and how one comes to the Shannon entropy to measure the uncertainty in information theory, we briefly recall the argument put forward in Ref.~\citet{shannon1948} to derive the definition of entropy as the unique function satisfting three necessary conditions.
\\
First, let us highlight that Shannon set the problem of finding a measure for the information as follows \citet{shannon1948}:
\\
\\
"\textit{Suppose we have a set of possible events whose probabilities of occurrence are $p_1, p_2, ..., p_n$. These probabilities are known but that is all we know concerning which event will occur. Can we find a measure of how much “choice” is involved in the selection of the event or of how uncertain we are of the outcome?}"
\\
\\
Shannon names this measure $H(p_1, p_2, ..., p_n)$, with normalized probabilities $\sum_np_n=1$, and defines it by requiring the following constraints.
\begin{enumerate}
\item \textit{Continuity}: $H$ should be continuous in $p_1, p_2, ..., p_n$. \label{cr1}
\item \textit{Equiprobable events}: if $p_1 = p_2 = ... = p_n$, then $H$ should be monotonically increasing with $n$. \label{cr2}
\item \textit{Propagation of uncertainty}: if there is a convex subset of $\mathbf{p}\equiv\{p_1, p_2, ..., p_n\}$ with linearly dependent probabilities, then $H(\mathbf{p})$ should be given by the sum of entropies for independent sets of events, namely $$H[tp_1,(1-t)p_1,p_3,...p_n] = H(p_1,p_3,...,p_n)+p_1H(t,1-t),$$ with $t\in[0,1]$. \label{cr3}
\end{enumerate}
These prescriptions identify the expression of $H(\mathbf{p})$.
\begin{theorem}[Shannon entropy]
The unique function $H$ satisfying simultaneously the constraints~\ref{cr1},~\ref{cr2} and~\ref{cr3} is given by
\begin{equation}
H(\mathbf{p}) = -K\sum_{n=1}^Np_n\log p_n
\end{equation}
where $K$ is a positive real constant.
\end{theorem}
Conventionally, the constant $K$ is set to one and the base of the logarithm is set to two. Moreover, here we will refer the entropy to a pertaining random variable $A$ indicating the addressed physical system, with corresponding probability distribution $\mathbf{p}_A\equiv\{p_A(a)\}_{a\in \mathcal{A}}$ with the discrete outcomes $a$ belonging to the set $\mathcal{A}$. Therefore, hereon the Shannon entropy will read
\begin{equation} \label{shentropy}
H(A) \equiv -\sum_a p_A(a)\log_2p_A(a) \quad \text{with},
\end{equation}
where, by convention (using the continuity condition), we set $0\log_20=0$.
If the random variable $A$ is dichotomic, then it is simple to see that the Shannon entropy reduces to
\begin{equation}\label{binaryentropy}
    h(A)=-p_A\log p_A-(1-p_A)\log(1-p_A),
\end{equation}
which is called \textit{binary entropy}.\\
\\
The Shannon entropy has been later extended to the case of continuous probability distribution.
\begin{definition}[Differential Shannon entropy]
Given a random variable $X$ with values $x\in \mathcal{X}$ and continuous probability distribution $\rm{Pr}_X(x)$, the differential,or continuous, Shannon entropy is defined as
\begin{equation} \label{contsh}
\Tilde{H}(X) \equiv -\int_{\mathcal{X}}dx\,{\rm Pr}_X(x)\log_2{\rm Pr}_X(x).
\end{equation}
\end{definition}
In contrast to the discrete case, this is a dimensionful quantity and is defined modulo an arbitrary additive value (as can be seen with a change of variables $x\rightarrow ax$). Here, we will mainly deal with discrete variables, and use Eq.~(\ref{shentropy}).
Now, starting from the definition of Shannon entropy, we introduce a crucial quantifier of the uncertainty which takes into account the a-priori knowledge of the observer about the system, namely the \textit{conditional entropy}. First, we remark that the Shannon entropy in Eq.~(\ref{shentropy}) can be used to evaluate the uncertainty of more than one random variable, say $A$ and $B$, by using the pertaining joint distribution $p_{AB}(a,b) = p_{AB}(A=a,B=b)$, yielding
\begin{equation}
    H(A,B) = - \sum_{a,b}p_{AB}(a,b)\log_2 p_{AB}(a,b),
\end{equation}
which defines the \textit{joint entropy}. This quantity is a measure of the global information hold by the systems $A$ and $B$, including the shared information. The difference between the joint entropy and the Shannon entropy of a single system determines the uncertainty about a system given the knowledge of the other.
\begin{definition}[Conditional entropy]
The entropy of a system $A$ conditioned on a system $B$ is given by
\begin{align} \label{condsh}
    H(A|B)&\equiv H(A,B)-H(B) \\
&=  - \sum_{a,b}p_{AB}(a,b)\log_2 p_{AB}(a|b), \nonumber
\end{align}
where $p_{AB}(a|b)$ is the conditional probability of $A=a$ given that $B=b$. 
\end{definition}
We derived the notion of conditional entropy from the Shannon entropy, but note that, actually, the former dwells at a more fundamental level: unconditional entropies can always be understood as the special cases of conditional entropies where the side information is uncorrelated with the observed system.
\\
An important property of the Shannon entropy that we will use is the \textit{subadditivity}, i.e.
\begin{equation}
H(A,B) \leq H(A) + H(B)
\end{equation}
where the equality holds only if the random variables $A$ and $B$ are independent. By using the definition of conditional entropy in Eq.~(\ref{condsh}), it is simple to see that the subadditivity property implies that \textit{conditioning never increases the entropy}, namely $H(A|B) \leq H(A)$.


\subsubsection{Rényi entropies}
\label{subsubsec312}
The constraint~\ref{cr3} can be relaxed to a much weaker condition, which, however, still yields a fundamental property of the entropy, and reads \citep{renyi1960}
\begin{itemize}
 \item \textit{Additivity}: if two events described by the distributions $\mathbf{p}\equiv \{p_1, p_2, ..., p_n\}$ and $\mathbf{q}\equiv \{q_1, q_2, ..., q_n\}$ are independent, then the entropy of the joint distribution is the sum of the entropies of the single events, i.e. $H(\mathbf{p}\times\mathbf{q}) = H(\mathbf{p}) + H(\mathbf{q})$, where $\times$ is the direct product. \label{cr4}
\end{itemize}
Of course, by relaxing the constraint~\ref{cr3} we cannot restrict to an unique function; on the contrary, the criteria~\ref{cr1} and \ref{cr2} with the additivity identify a one-parameter family of quantities, known as \textit{Rényi entropies}. These functions can be defined as the self-information of the $\alpha$-norm of a vector of probabilities $\mathbf{p}_A$ \citep{wehner2010}.
\begin{definition}[Rényi entropies]
Given the vector of probabilities $\mathbf{p}_A$ with
\begin{equation}
\lVert\mathbf{p}_A\rVert_{\alpha} = \left(\sum_{a\in \mathcal{A}}p_A(a)^{\alpha}\right)^{1/\alpha},
\end{equation}
the Rényi entropy of $\mathbf{p}_A$ is defined as
\begin{equation}
H_{\alpha}(A)= \frac{\alpha}{1-\alpha}\log_2\lVert\mathbf{p}_A\rVert_{\alpha}
\end{equation}
for any $\alpha\geq 0$.
\end{definition}
In the limit $\alpha\rightarrow1$, the Shannon entropy is re-obtained. Other relevant cases are the \textit{Hartley entropy}
\begin{equation} \label{coll}
H_0(A) = -\log_2|\mathcal{A}| \equiv H_{\text{max}}(A),
\end{equation}
the \textit{collision entropy}
\begin{equation} \label{coll}
H_2(A) = -\log_2\sum_{a\in\mathcal{A}}p_A(a)^2,
\end{equation}
and the \textit{min-entropy}
\begin{equation} \label{min}
H_{\infty}(A) \equiv H_{\text{min}}(A) = -\log_2\max_{a\in\mathcal{A}}p_A(a).
\end{equation}
\\
The Rényi entropies are monotonically decreasing in the parameter $\alpha$, implying $H_{\infty}(A)\leq H_2(A)\leq H(A) \leq H_0(A)$. This generalization of the Shannon entropy has allowed to achieve a very rich characterization of the uncertainty in different random processes, as we will see in the following.

\subsubsection{Smooth Rényi entropies}
\label{subsubsec313}
Finally, we see an entropy measure specifically devised to address tasks in information theory. 
As such, curiously, it has been formerly conceptualized in the context of specific applications and later formalized in a more general scenario, which is the opposite process that led to define the Shannon and Rényi entropies. Indeed, the notion of \textit{smooth entropy} as an uncertainty measure has been introduced by \citet{cachin1997} to solve a problem at the interception of cryptography and theoretical computer science, namely finding a formulation that could unify \textit{privacy amplification} and \textit{entropy smoothing}. The former is the final step of a key distribution protocol, where a secret key is extracted by applying a random function on the strings of two or more parties, in the presence of an adversary holding partial knowledge of the information shared by the parties. Entropy smoothing is the conversion of an arbitrary random string into a string with smaller alphabet and where the bits are \textit{almost uniformly} distributed. Within this context, \textit{the smooth entropy is defined as the uncertainty measure that quantifies the number of almost uniformly distributed bits that a probabilistic algorithm can extract from an element of a set of random variables}. The distribution of the output bits is \textit{almost} uniform because the smooth entropy allows to consider small correlations with the random bits employed in the smoothing process, thus accounting for arbitrarily small deviations from the uniform randomness of the output bits. As such, it is an operational measure that cannot be computed directly from a given distribution, differently from the Shannon and Rényi entropies, but it requires to define a notion of distance in the probability space. Importantly, we remark that the smooth entropy is the smoothed version of a specific entropy. Therefore, it can be defined after having devised how the definition of entropy smoothing can be applied to the entropy to be generalized.
\\
Connections between the smooth and Rényi entropies have been first inspected by \citet{cachin1997b}, but it is with \citet{renner2004} that the smooth entropy has been formalized as an extension of Rényi entropies, named \textit{smooth Rényi entropies}.
They quantify the statistical uncertainty of a random variable in a twofold way: in the sense of a specific $\alpha$-entropy and in terms of the distance $\Delta$ between the distribution $\{p(a)\}_{a \in \mathcal{A}}$ of the random variable and another distribution $\{q(a)\}_{a \in \mathcal{A}}$ over the same set $\mathcal{A}$
\begin{equation} \label{dist1}
\Delta\left(\{p(a)\}_{a \in \mathcal{A}}, \{q(a)\}_{a \in \mathcal{A}}\right) \equiv \frac{1}{2}\sum_{a\in\mathcal{A}}\left|p(a) - q(a)\right|.
\end{equation}
\begin{definition}[Smooth Rényi entropy]
Given a random variable $A$ with probability distribution $\mathcal{P}_{\mathcal{A}}\equiv\{p(a)\}_{a \in \mathcal{A}}$ within the set $\mathcal{A}$, and the parameters $\alpha,\varepsilon \geq 0$, the $\varepsilon$-smooth Rényi entropy of $A$ is defined as
\begin{equation}
H_{\alpha}^{\varepsilon}(A) \equiv \frac{1}{1-\alpha}\inf_{\mathcal{Q}_{\mathcal{A}}\in \mathcal{C}^{\varepsilon}(\mathcal{P}_{\mathcal{A}})}\log_2\left(\sum_{a\in\mathcal{A}}q(a)^{\alpha}\right)
\end{equation}
where 
\begin{equation}
\mathcal{C}^{\varepsilon}(\mathcal{P}_{\mathcal{A}}) \equiv \{\mathcal{Q}_{\mathcal{A}} : \Delta\left(\mathcal{P}_{\mathcal{A}},\mathcal{Q}_{\mathcal{A}}\right)\leq\varepsilon\}
\end{equation}
is the set of probability distributions $\mathcal{Q}_{\mathcal{A}}\equiv\{q(a)\}_{a \in \mathcal{A}}$ which are $\varepsilon$-close to $\mathcal{P}_{\mathcal{A}}$, in terms of the distance in Eq.~(\ref{dist1}).
\end{definition}

Like the Rényi entropies, this measure is monotonically decreasing for increasing $\alpha$, namely
\begin{equation}
H_{\alpha}^{\varepsilon}(A) \geq H_{\beta}^{\varepsilon}(A) \quad \text{for} \quad \alpha \leq \beta.
\end{equation}
Crucially, the smooth Rényi entropies featuere two key properties that define their operational meaning. We report them in the following two lemmas.
\begin{lemma} \label{lemma1}
Consider an $n$-tuple $A^n=(A_1, ..., A_n)$ of independent random variables $A_j$, each of them sharing the same distribution $\mathcal{P}_{\mathcal{A}}\equiv\{p(a)\}_{a \in \mathcal{A}}$ of a random variable $A$. Then,
\begin{equation}
\lim_{\varepsilon\to 0}\lim_{n\to \infty}\frac{H_{\alpha}^{\varepsilon}(A^n)}{n} = H(A) \quad \forall\,\alpha\geq 0.
\end{equation}
\end{lemma}
\begin{lemma} \label{lemma2}
Given a random variable $A$ and the parameters $\varepsilon > 0$, $\alpha\geq 0$,
\begin{align}
&H_{\text{min}}^{\varepsilon}(A) \geq H_{\alpha}(A) - \frac{\log_2\varepsilon}{\alpha -1} \quad \forall\,\alpha >1 \\ 
&H_{\text{max}}^{\varepsilon}(A) \leq H_{\alpha}(A) + \frac{\log_2\varepsilon}{1 -\alpha} \quad \forall\,\alpha <1.
\end{align}
\end{lemma}
Lemma~\ref{lemma1} establishes the convergence of the smooth Rényi entropy to the Shannon entropy for every order $\alpha$ \textit{in the case of a large number of independent random variables}. This property allows to compute some essential quantities in information theory in scenarios where the Shannon entropy fail to identify them. A remarkable example is the problem of data compression \citep{konig2009}. The \textit{compression rate} $r^{\varepsilon}(A^n)$ of a source emitting a sequence $A^n$ of symbols $A_j$, with $j \in [1,n]$, each of them distributed with probability $\mathcal{P}_{\mathcal{A}}$, is the minimum fraction of the $n$ bits needed to encode $A^n$ such that $A^n$ can be recovered from the encoding with error probability up to $\varepsilon$. From the \textit{source-coding theorem} \citep{shannon1948}, we know that, in the limit of \textit{asymptotically large} $n$, \textit{small} $\varepsilon$ and for independent $A_j$ (or, also, for Markovian $A^n$), the compression rate is approximated by the Shannon entropy of a random variable $A$ distributed as the symbols $A_j$, namely
\begin{equation} \label{sct}
\lim_{\varepsilon\to 0}\lim_{n\to \infty}r^{\varepsilon}(A^n) = H(A).
\end{equation}
While this provides a good estimation in many cases of interest, there are practical scenarios where some or all of the hypothesis under Eq.~(\ref{sct}) are not verified, especially in cryptography and randomness extraction. For instance, this is the case if one has to deal with a finite number of resources or if there are correlations among the symbols of the string. In such circumstances, the Shannon entropy must be replaced by a suitable smooth Rényi entropy \citep{konig2009}.
\\
But which of the different smooth Rényi entropies labeled by the parameter $\alpha$ shall be choosed? Lemma~\ref{lemma2} shows that the Rényi entropies are bounded by one of the two extremal, with respect to $\alpha$, smooth entropies, $H^{\varepsilon}_{\text{min}}$ and $H^{\varepsilon}_{\text{max}}$, for every order $\alpha$. This result can be used to show \citep{renner2004} that many key properties of a random variable can be fully characterized by these two quantities. We will inspect some of them in the quantum framework, where the smooth min- and max-entropy take a distinct physical meaning in terms of uncertainty relations.

\subsubsection{Extension to quantum information theory}
\label{subsubsec314}
When addressing a quantum system, we need to consider that the probability distributions defining the entropies reviewed so far are obtained by measuring an observable $\hat{A}$ on a state $\rho$. Therefore, we write $H_{\alpha}(\hat{A})_{\rho}$, with the probabilities obtained from the Born rule, namely
\begin{equation}
p_A(a|\rho) = \text{Tr}[\rho\, \Pi_{A}^{(a)}],
\end{equation}
where $\Pi_{A}^{(a)}$ is either a projection or, more generally, a positive-operator valued measurement (POVM) on $A$ yielding the outcome $a$.
In most of the cases, it will not be necessary to specify the state, so we will omit it and write $H_{\alpha}(\hat{A})$.
\\
However, in quantum mechanics the state itself is described by an operator, i.e. the density matrix $\rho$, describing statistical mixtures of pure states. As such, it features an intrinsic statistical uncertainty, that can be described by a specific entropy \citep{vonneumann1932}.
\begin{definition}[von Neumann entropy]
Given a Hilbert space $\mathcal{H}$, the entropy of a state 
\begin{equation}
\rho = \sum_{x}\lambda_x|\lambda_x\rangle\langle\lambda_x| \in \mathcal{S}(\mathcal{H})
\end{equation}
with
\begin{equation}
\mathcal{S}(\mathcal{H}) \equiv \{\rho \in \mathcal{H} | \rho \geq 0, \rm{Tr}[\rho] = 1\},
\end{equation}
is defined as
\begin{equation}
S(\rho)\equiv -\mathrm{Tr}(\rho\log_2\rho) = -\sum_{x}\lambda_{x}\log_2\lambda_{x}.
\end{equation}
\end{definition}
Note that, by definition,
\begin{equation}\label{purevsmixed}
   0\leq S(\rho)\leq\log_2 d\quad \text{where}\quad\begin{cases}
    S(\rho)=0\quad \text{for pure states}\\
    S(\rho)=\log_2d\quad \text{for maximally mixed states}.
\end{cases}
\end{equation}

As for the Shannon entropies, given a composite system $AB$ with density matrix $\rho_{AB}$, we can define the joint and the conditional Von Neumann entropy as
\begin{align}
    &S(\rho_{AB})\equiv -\text{Tr}\left[\rho_{AB}\log_2(\rho_{AB})\right]\\
    &S(\rho_A|\rho_B)\equiv S(\rho_{AB})-S(\rho_{B}),
\end{align}
satisfying the subadditivity, i.e. $S(\rho_{AB}) \leq S(\rho_A)+ S(\rho_B)$. In the case of a tripartite system $\rho_{ABC}$, the von Neumann entropy satisfies the \textit{quantum strong subadditivity}, reading
\begin{equation}
S(\rho_{ABC}) + S(\rho_B) \geq S(\rho_{AB}) + S(\rho_{BC}),
\end{equation}
whose proof has been provided by \citet{lieb1973}.
Among the basic properties of the von Neumann entropy, we will need the one that allows to know how to compute it for a bipartite state with classical correlations and classical side information \citep{nielsen2010}.
\begin{theorem}[Joint Entropy Theorem] \label{JET}
Given an observable $\hat{A}$ with eigenstates $\{|a\rangle\langle a|\}_{a \in \mathcal{A}}$, a probability distribution $p(a)$ on $\mathcal{A}$ and a set of states $\rho_{B|a}$ labeled by $a \in \mathcal{A}$, we have
\begin{equation}
S\left(\sum_{a\in\mathcal{A}}p(a)|a\rangle\langle a| \otimes \rho_{B|a}\right) = H\left(\{p(a)\}_{a\in\mathcal{A}}\right) + \sum_{a\in\mathcal{A}}p(a)S(\rho_{B|a}).
\end{equation}
\end{theorem}
Theorem~\ref{JET} establishes two crucial points:
\begin{enumerate}
\item  the von Neumann entropy of a product state $\sigma_A\otimes\rho_B$ is the sum of the entropies of the states,
\item the von Neumann entropy of a statistical mixture of orthonormal states is the Shannon entropy of the pertaining distribution.
\end{enumerate}
A striking feature of the conditional von Neumann entropies is that they can be negative in the presence of quantum correlations \citep{devetak2005}. Indeed, if a bipartite state $\rho_{AB}$ is entangled, then the information encoded on $\rho_{AB}$ cannot be reconstructed from $\rho_A$ and/or $\rho_B$ separately, so that $S(\rho_{AB}) \leq S(\rho_B)$ and $S(\rho_{AB}) \leq S(\rho_A)$. Therefore, negative conditional entropies are a signature of quantum correlations.

The careful reader has probably noted that we have not defined a notion of conditional entropy in the case of Rényi and smooth entropies, although we have highlighted their essential role in section~\ref{subsubsec311}. Indeed, conditioning on these entropies is not straightforward as for the Shannon entropies and has required long-lasting debates and research. In particular, the derivation of the conditional Rényi entropy involves sophisticated technicalities that are beyond the scope of this review. For the interested reader, we suggest Refs.~\citep{arimoto1977,fehr2014,sason2017} for the classical foundation and Refs.~\citep{konig2009,datta2009,muller2013,tomamichel2015} for the quantum case. Here we present the definitions of the conditional Rényi entropies and their smoothed generalization as they are used in quantum information. We will see in sections~\ref{sec4} and~\ref{sec5} that different linear combinations of these quantities are bounded by specific URs that are crucial in many applications.
\begin{definition}[Sandwiched quantum Rényi entropies] \label{sqre}
For any state $\rho_{AB}$, given a parameter $\alpha \geq 1/2$, the sandwiched Rényi entropy of a system $A$ conditioned on $B$ is defined as
\begin{equation} \label{sand}
H_{\alpha}(A|B) \equiv \sup_{\sigma_B\in\mathcal{S}(\mathcal{H}_B)}\frac{1}{1-\alpha}\log_2\rm{Tr}\left[\left(\mathbb{I}_A\otimes\sigma_B^{\frac{1-\alpha}{2\alpha}}\rho_{AB}\,\mathbb{I}_A\otimes\sigma_B^{\frac{1-\alpha}{2\alpha}}\right)^{\alpha}\right],
\end{equation}
where $\mathcal{H}_B$ is the Hilbert space of system $B$.
\end{definition}
Note that, differently from the unconditional classical Rényi entropies defined in section~\ref{subsubsec312}, in this formulation the entropies are not defined for $\alpha < 1/2$. This is due to the axiomatic construction \citep{muller2013,tomamichel2015} of the sandwiched Rényi entropies from a related quantity, i.e. the Rényi relative entropy: in the case $\alpha < 1/2$ one fo the fundamental conditions required for the definition of the relative entropy is not satisfied.
\\
Like the Rényi entropies, the sandwiched entropies are monotonically decreasing in $\alpha$: $H_{\alpha}(A|B) \geq H_{\beta}(A|B)$ for $\alpha \leq \beta$. Therefore, in this case the max-entropy is identified by $\alpha = 1/2$, namely $H_{\text{max}}(A|B) \equiv H_{1/2}(A|B)$. Here we will mainly focus on the exrtremal Rényi entropies, i.e. the min- and max- entropies, since, as for the classical case (see section~\ref{subsubsec313}), when generalized to their smooth version, they describe the main quantities of interest in quantum information, and asymptotically yield the conditional Shannon entropy.
Moreover, the min-and max-entropy can be formulated as semi-definite programs \citep{tomamichel2015}, which is an operational approach that helps to handle their optimization. In particular, the conditional min-entropy can be equivalently defined as
\begin{equation} \label{sdpmin}
H_{\text{min}}(A|B) = \sup\left[\lambda : \exists\sigma_B\in\mathcal{S}(\mathcal{H}_B) \wedge \rho_{AE} \leq 2^{-\lambda}\,\mathbb{I}_A\otimes\sigma_B \right].
\end{equation}

Starting from the construction of the conditional Rényi entropies as sandwiched entropies in Eq.~(\ref{sand}), we can finally extend also the smooth min- and max- entropies to conditional quantities in quantum information. If for the classical case introduced in section~\ref{subsubsec313} we defined a distance between probability distributions to identify the ball of radius $\varepsilon$ [see Eq.~(\ref{dist1})], similarly here we will need a notion of distance between states.


\begin{definition}[Purified distance] \label{purdi}
The purified distance between two quantum states $\rho$ and $\sigma$ is defined as
\begin{equation}
    D_{P}(\rho,\sigma) \equiv\sqrt{1-F(\rho,\sigma)},
\end{equation}
where $F$ is the generalized fidelity
\begin{equation}
    F(\rho,\sigma) \equiv \left[\rm{Tr}\sqrt{\sqrt{\rho}\sigma\sqrt{\rho}} + \sqrt{(1-\rm{Tr}\rho)(1-\rm{Tr}\sigma)}\right].
\end{equation}
\end{definition}
We have mentioned that it is useful to re-formulate the min- and max- entropy, as in Eq.~(\ref{sdpmin}), in terms of semi-definite programming. One of the main applications of this approach is the very definition of their smooth version. Indeed, the smooth min- and max- entropies are the optimization of the corresponding sandwiched Rényi entropies over states that are $\varepsilon$-close in the purified distance.
\begin{definition}[Quantum smooth min- and max-entropies]
Given a state $\rho_{AB}$, the $\epsilon$-smooth min-entropy and the $\epsilon$-smooth max-entropy are defined as
\begin{equation}
    H^{\varepsilon}_{\rm{min}}(A|B)_{\rho}\equiv\max_{\tilde{\rho}\in \mathcal{C}^{\epsilon}(\rho)}H_{\rm{min}}(A|B)_{\tilde{\rho}},
\end{equation}
and
\begin{equation}
    H^{\varepsilon}_{\rm{max}}(A|B)_{\rho} \equiv \min_{\tilde{\rho}\in \mathcal{C}^{\epsilon}(\rho)}H_{\rm{max}}(A|B)_{\tilde{\rho}},
\end{equation}
where $\mathcal{C}^{\varepsilon}(\rho)$ is the ball centered in $\rho$ with radius $\varepsilon$ such that
\begin{equation}
    \mathcal{C}^{\varepsilon}(\rho)=\left\{\tilde{\rho}_{AB}\in \mathcal{S}(\mathcal{H}_AB):\tilde{\rho}_{AB}\geq 0,\,{\rm Tr}(\tilde{\rho}_{AB})\leq 1,\, D_{P}(\rho_{AB},\tilde{\rho}_{AB})\leq \varepsilon \right\}.
\end{equation}
\end{definition}
Crucially, similarly to the classical case, in the limit of several copies of a state, the smooth min- and max-entropy converge to the von Neumann entropy, i.e.
\begin{equation}
\lim_{n\to\infty}\frac{1}{n}H^{\varepsilon}_{\text{min}}(A^n|B^n)_{\rho^{\otimes n}} = \lim_{n\to\infty}\frac{1}{n}H^{\varepsilon}_{\text{max}}(A^n|B^n)_{\rho^{\otimes n}} = H(A|B)_{\rho}.
\end{equation}
Therefore, if we are characterizing a resource usage in the one-shot setting, the rate of recource usage in the limit of independent, identically distributed and many repetitions is given by the von Neumann entropy.
\\
Conversely, the operational meaning of the smooth min- and max-entropy is defined in the case of a finite amount of resources. The conditional smooth min-entropy $H^{\varepsilon}_{\text{min}}(A|B)$ quantifies the number of bits of system $A$ that are $\varepsilon$ close to a string which is uniformly distributed and independent with respect to system $B$. In other words, the number of bits of $A$ that $B$ can guess with probability less than $\varepsilon$. The conditional smooth max-entropy $H^{\varepsilon}_{\rm{max}}(A|B)$ quantifies the number of bits that one needs to reconstruct $A$ from the information of $B$ up to a failure probability $\varepsilon$.
%
\section{The entropic formulation of the uncertainty relations}
\label{sec4} 
\subsection{The Entropic Uncertainty Relations, from Hirschman to Maassen - Uffink}
\label{subsec31}
We have seen different notions of entropy that capture a rich description of the concept of statistical uncertainty. We introduce the \textit{entropic uncertainty relations (EURs)}, namely relations that use the entropy to assess the complementarity of quantum observables. In particular, in this section we will start by entropies \textit{without} side information and will let the observer's knowledge come into play in section~\ref{sec5}.
\\
The first known attempt in this direction has been the one by \citet{hirschmann1957}, who introduced an EUR for a pair of continuous variables related by Fourier transform, such as position and momentum, in terms of the differential Shannon entropy, introduced in Eq.~(\ref{contsh}).
Later, \citet{beckner1975} improved his inequality and finally \citet{bialynicki1975} proved the EUR
\begin{equation} \label{birula}
\tilde{H}(\hat{Q}) + \tilde{H}(\hat{P}) \geq D(1+\log_2\pi)
\end{equation}
which they proved to be stronger than the Heisenberg UR, namely  Eq.~(\ref{birula}) implies Eq.~(\ref{heisenberg}) in dimension $D$. In particular, here $D$ is the dimension of the position and momentum space.
\\
The intuition that established the EURs as a fundamental formulation of the uncertainty principle and an essential tool for many tasks in quantum information theory came with \citet{deutsch1983}, who first realized the weakness of the Robertson (and Schr\"odinger) UR and understood that it could be overcome by formulating a suitable EUR. As discussed above, the lower bound is state dependent and becomes trivial for states giving zero mean value of the commutator and anticommutator equal to the product of the expectation values, which in the case of finite dimension and discrete variables is always possible. \citet{deutsch1983} devised the first EUR for discrete variables with the Shannon entropies defined in Eq.~(\ref{shentropy}) for an arbitrary pair of observables $\hat{A}$ and $\hat{B}$, yielding a state-independent bound, namely
\begin{equation} \label{deutsch}
H(\hat{A}) + H(\hat{B}) \geq -2\log_2\left(\frac{1+\sqrt{c}}{2}\right),
\end{equation}
where the measurements of the observables $\hat{A}$ and $\hat{B}$ are projective, $\Pi_A^{(a)}=\{|a\rangle\langle a|\}_{a\in \mathcal{A}}$ and $\Pi_B^{(b)}=\{|b\rangle\langle b|\}_{b\in \mathcal{B}}$, and the \textit{incompatibility factor} $c$ is defined as
\begin{equation} \label{incompa}
c \equiv c\left(\Pi_A,\Pi_B\right) = \max_{a\in \mathcal{A},b\in \mathcal{B}}\left[|\langle a| b\rangle|^2\right]
\end{equation}
and identifies the maximum overlap between the two measurements. Therefore, we see that the EURs set a bound on the knowledge of complementary observables which \textit{depends uniquely on the choice of the observables}. Crucially, the sum of the entropies [left-hand side of Eq.~(\ref{deutsch})] depends on the state preparation, but the lower bound [right-hand side of Eq.~(\ref{deutsch})], which is the ultimate limit on the information that can be extracted from their measurements, does not, thus determining the origin of quantum uncertainty in the very structure of quantum observables and measurements, and independently of the state of the system.
\\
By proving and extending a conjecture by \citet{kraus1987}, this EUR was later strenghtened and generalized in terms of Rényi entropies by \citet{maassen1988}.
\begin{theorem}[Maassen-Uffink generalized EUR] \label{thmu}
Given a state $\rho \in \mathcal{H}$ with $\text{dim}\left(\mathcal{H}\right) = d$ and a pair of observables $\hat{A}$ and $\hat{B}$ with orthonormal bases $\{|a_j\rangle\}_{j=1}^d$ and $\{|b_k\rangle\}_{k=1}^d$, respectively, then
\begin{equation} \label{mueur}
H_{\alpha}(\hat{A}) + H_{\beta}(\hat{B}) \geq -\log_2c
\end{equation}
with $c$ as defined in Eq.~(\ref{incompa}), and $\alpha$, $\beta$ such that $\alpha^{-1}+\beta^{-1}=2$.
\end{theorem}
Kraus' conjecture is the case $\alpha, \beta \rightarrow 1$, i.e. the Maassen-Uffink EUR in terms of Shannon entropies, stronger than the Deutsch relation, in Eq.~(\ref{deutsch}). The proof of Eq.~(\ref{mueur}) was obtained by exploiting a representation theorem for $\alpha$-norms by \citet{riesz1929}, but other derivations are known \citep{coles2011,coles2017}. A relevant specific case of Eq.~(\ref{mueur}) is the EUR that connects min and max entropy, namely
\begin{equation}
H_{\text{min}}(\hat{A}) + H_{\text{max}}(\hat{B}) \geq -\log_2c.
\end{equation}
Its generalization to the case with side information and a smoothing parameter defines the security of quantum secrecy systems, due to the operational meaning of the min entropy. Interestingly, in the same work \citep{maassen1988}, the authors derived a relation for the min entropies analog to the result by \citet{deutsch1983}, reading
\begin{equation} \label{minmin}
H_{\text{min}}(\hat{A}) + H_{\text{min}}(\hat{B}) \geq -2\log_2\left(\frac{1+\sqrt{c}}{2}\right),
\end{equation}
which extends a previous result by \citet{landau1961}.
\\
Let us focus on the lower bound, again determined uniquely by the incompatibility factor. As expected from the notion of complementarity, it annihilates if the observables share at least one common eigenvector.
Conversely, the bound is maximum if the bases of the observables are \textit{mutually unbiased}.
\begin{definition}[Mutual unbiasedness]
Given two bases, $\{|a_j\rangle\}_{j=1}^d$ and $\{|b_k\rangle\}_{k=1}^d$, belonging to a Hilbert space $\mathcal{H}$ with $\text{dim}\left(\mathcal{H}\right) = d$, we say that they are mutually unbiased if measuring one with the projectors of the other provides a uniform distribution of the outcomes, namely
\begin{equation} \label{mubs}
|\langle a_j|b_k\rangle| = \frac{1}{\sqrt{d}} \quad \forall\,j,k\in[1,d].
\end{equation}
\end{definition}
Therefore, if the observables are mutually unbiased, the lower bound in Eq.~(\ref{mueur}) is given by $\log_2d$, and, importantly, \textit{it is the largest tight bound that can be obtained for any choice of the two observables and for any state}. In this sense, we say that the Maassen-Uffink EURs with mutually unbiased bases are \textit{maximally strong} \citep{wehner2010}. Interestingly, here we see how the EURs capture crucial complementary properties of quantum observables. We will see in section~\ref{subsec33} that the EURs can be extended to more than two observables and will find that, in those cases, there is not a general result, such as Theorem~\ref{thmu}, assessing the dependence of the lower bound on the spectral properties of the observables. The set of observables leading to maximally strong EURs probably is not a set of mutually unbiased bases, as defined by Eq.~(\ref{mueur}). The conditions that a set of observables should fulfill to provide maximally strong EURs has been proposed by \citep{wehner2010} and define a \textit{maximally incompatible set}.

\begin{definition}[Maximally incompatible set] \citep{wehner2010} \label{maxinc}
Given a set of $m$ $d$-dimensional observables $\{\hat{A}_j\}_{j=1}^m$, we say that the pertaining set of projections $\{\Pi_{A_j}\}_{j=1}^m$, with $\Pi_{A_j}=\{|a_j^{(k)}\rangle\langle a_j^{(k)|}\}_{k=1}^d$, is maximally incompatible if 
\begin{itemize}
\item it is always possible to find a state $\rho$ such that $H(\hat{A}_{\bar{j}}) = 0$, with $\bar{j} \in [1,m]$ and $H(\hat{A}_j)\neq 0 \, \forall\,j\neq\bar{j}$,
\item such state $\rho$ is the one saturating the corresponding EUR, which is, therefore, tight.
\end{itemize}
\end{definition}
Note that, in general, it is not guaranteed that such a set exists for every dimension and number of observables. Mutually unbiased bases are maximally incompatible in the case with $m = 2$ and generic dimension, but they are not for $m > 2$ and few is still known in this regime.

\subsection{Tighter bounds for pairs of observables}
\label{subsec32}
Some attempts have been made to improve the Maassen-Uffink EUR in terms of the Shannon entropies, by leveraging on the properties of  specific systems and thus finding the largest function of $c$ bounding the sum of the entropies. We remark that, by claiming that an EUR is \textit{improved}, we mean that it is stronger in the sense of Definition~\ref{maxinc}: the bound is tighter.
\\
In the case of qubits, \citet{sanchez1998} obtained
\begin{equation}
H(\hat{A}) + H(\hat{B}) \geq h\left(\frac{1+\sqrt{2c-1}}{2}\right),
\end{equation}
where $h(\cdot)$ is the binary entropy defined in Eq.~(\ref{binaryentropy}). Subsequently, \citet{ghirardi2003} managed to recast the search for the tightest bound for qubit systems into an optimization problem, for which they found an analytical solution given $c \geq 1/\sqrt{2}$, reading
\begin{equation} \label{gmr}
H(\hat{A}) + H(\hat{B}) \geq 2h\left(\frac{1+\sqrt{c}}{2}\right).
\end{equation}

\citet{devicente2008} showed that the EUR in Eq.~(\ref{gmr}) holds for a generic finite dimension $d\geq 2$, with the same constraint on the incompatibility factor. That bound was further strengthened by exploiting the second-largest contribution $c_2$ of the overlap matrix $|\langle a_j|b_k\rangle|^2$. In this direction, two similar but distinct results have been obtained: the Coles-Piani EUR \citep{coles2014}
\begin{equation} \label{cp}
H(\hat{A}) + H(\hat{B}) \geq \frac{1-\sqrt{c}}{2}\log_2\frac{c}{c_2} - \log_2c
\end{equation}

and the Rudnicki-Pucha{\l}a-\.Zyczkowski EUR \citep{rudnicki2014}
\begin{equation} \label{rpz}
H(\hat{A}) + H(\hat{B}) \geq -\log_2\left[\left(\frac{1+\sqrt{c}}{2}\right)^2(c-c_2) +c_2\right].
\end{equation}

Applications to the more general case of Rényi entropies has been explored in Ref.~\citep{abdelkhalek2015}, where the uncertainty of pairs of complementary observables has been fully characterized.
\\
Specific EURs for spin observables, stronger than the EURs in Eqs.~(\ref{cp}) and~(\ref{rpz}) has been obtained for the cases of spin $1$, $3/2$ and $2$ by \citet{riccardi2017}.



\subsection{Extensions to more than two observables}
\label{subsec33}
Here, we present the main extensions of the EURs to more than two observables. In this framework, the case of mutually unbiased bases has been widely addressed, since they feature the strongest bound in the Maassen-Uffink EUR. In prime power dimensions, there are $d+1$ mutually unbiased bases, while, for all the other cases, at least three are known. 
The explicit construction of a complete set of mutually unbiased bases is in general a hard problem, which has been solved only for specific cases. There are at least three methods to address this task: by means of the Pauli-Heisenberg operators, via representation with Hadamard matrices or by using Galois fields. We briefly review in~\ref{app1} and~\ref{app2} the first two approaches. We will mainly need the second one.
\\
\\
We will focus on the EURs in terms of Shannon entropies. Given a set of $m$ observables $\{\hat{A}_j\}_{j= 1}^{m}$ in dimension $d$ with mutually unbiased eigenstates, the general expression of the addressed inequalities reads
\begin{equation} \label{geneur}
    \sum_{j=1}^{m}H(\hat{A}_j) \geq \mathcal{B}_{d,m}\,,
\end{equation}
namely, the lower bound $\mathcal{B}$ on the sum of the Shannon entropies of the observables is specific for each $d$ and $m$ considered. Firstly, we note that just iterating the Maassen-Uffink EUR for all the $m$ observables,one gets the trivial bound
\begin{equation} \label{trivial}
 \mathcal{B}_{d,m} = \frac{m}{2}\log_2d,
\end{equation}
which is, indeed, a weak EUR. However, unexpectedly, it is tight for some sets of mutually unbiased bases. It has been proved by \cite{ballester2007} that one can derive, from the Pauli-Heisenberg operators approach in~\ref{app2}, up to $m=p^l+1$ mutually unbiased bases with the sum of entropies constrained by the bound in Eq.~(\ref{trivial}), given that the dimension is a square prime power, $d=p^{2l}$ with $p$ prime and $l\in \mathbb{N}$. More cases where mutually unbiased bases fail to fulfil the requirements to be a maximally incomaptible set will be shown in the following. We can conclude that \textit{mutual unbiasedness is not a sufficient condition to find maximally strong EUR}. 
\\
\\
Two EURs have been assessed in terms of the collision entropy, defined in Eq.~(\ref{coll}), which, due to the monotonicity of the Rényi entropies, induce two corresponding EURs for the Shannon entropies. In the case of a complete set, with $m = d+1$, \citet{ivanovic1991} and \citet{sanchez1993} showed that
\begin{equation} \label{ivano}
\sum_{j=1}^{d+1}H_2(\hat{A}_j) \geq (d+1)\left[\log_2(d+1) -1\right],
\end{equation}
whose proof has been variously re-formulated \citep{ballester2007,wehner2010}. In particular, in Ref.~\citep{wehner2010}, an EUR for the case $m < d+1$ has been found and proved. It reads
\begin{equation} \label{stefi}
 \sum_{j=1}^{d+1}H_2(\hat{A}_j) \geq m\log_2\left(\frac{d\cdot m}{d+m-1}\right),
\end{equation}
and yields the bound in Eq.~(\ref{ivano}) for $m=d+1$.
The EUR in Eq.~(\ref{ivano}), expressed in terms of Shannon entropies, has been later improved by \citet{sanchez1995} in the specific case of even dimension, and reads
\begin{equation} \label{sancho}
 \mathcal{B}_{d,m} = \frac{d}{2}\log_2\frac{d}{2} + \frac{d+1}{2}\log_2\frac{d+1}{2}.
\end{equation}
These bounds, from Eqs.~(\ref{trivial}),~(\ref{ivano}),~(\ref{stefi}) and~(\ref{sancho}), have been numerically tested in dimensions three to five, for all number of bases, by \citet{riccardi2017}, and, when they did not prove to be tight, new EURs have been established. The caclculations were performed by exploiting the construction of the mutually unbiased bases via Hadamard matrices, detailed in~\ref{app1}. We report the bounds retrieved there in Table~\ref{bounds}. Simple calculations show that the bound in Eq.~(\ref{sancho}) is tight in dimension two, but this is never the case in dimension four. The trivial bound in Eq.~(\ref{trivial}), as expected, is always tight for $m=2$, but it is also consistent with $\mathcal{B}_{4,3}$, as this is the case where $d=p^{2l}$ and $m=p^l+1$, with $p=2$ and $l=1$, mentioned in the discussion after Eq.~(\ref{trivial}). The bound $\mathcal{B}_{4,3}$ is reproduced also by Eq.~(\ref{stefi}), while the bound in Eq.~(\ref{ivano}) is tight for $\mathcal{B}_{3,4}$.

Interestingly, the bound $\mathcal{B}_{5,3}$ is not unique. Indeed, it was found in Refs.~\citep{serino2025}, that it \textit{depends on the choice of the triple of bases considered in the uncertainty relation}, an effect that was named \textit{complementarity-based complementarity}. This is due to the existence, in dimension five, of two sets of triples of mutually unbiased bases which are inequivalent under unitary transformation, as proved by \citet{brierly2010}. In the first of the Refs.~\citep{serino2025}, it is demonstrated that this is a necessary condition for having two distinct EURs, one for each inequivalent set.

\begin{table}
\centering
\begin{tabular}{|c|c|c|c|c|c|} 
\cline{2-6}
\multicolumn{1}{c|}{} & $m=2$ & $m=3$ & $m=4$ & $\,m=5\,$ & $m=6$ \\
\cline{2-6}
\hline
$d=3$ & $\log_2 3$ & 3 & 4 & - & - \\
$d=4$ & 2 & 3 & 5 & 7 & - \\
$d=5$ & $\log_2 5$ & $4.43 \vee 2\log_2 5$ & 6.34 & 8.33 & 10.25 \\
\hline
\end{tabular}
\caption{Lower bounds $\mathcal{B}_{d,m}$ of the EURs in Eq.~(\ref{geneur}) for dimension $d$ three to five and $m$ MUBs.}
\label{bounds}
\end{table}
In the case of min-entropies, two extensions to Eq.~(\ref{minmin}) has been found for an arbitrary number of observables by \citet{mandayam2010}, reading

\begin{align}
&\sum_{j=1}^{m}H_{\text{min}}(\hat{A}_j) \geq -m\log_2\left[\frac{1}{m}\left(1+\frac{m-1}{\sqrt{d}}\right)\right] \\
&\sum_{j=1}^{m}H_{\text{min}}(\hat{A}_j) \geq -m\log_2\left[\frac{1}{d}\left(1+\frac{d-1}{\sqrt{m}}\right)\right],
\end{align}
each of them tight in different regimes.

\subsection{Extension to mixed states}
\label{subsec34}
The Maassen-Uffink EUR in Eq.~(\ref{mueur}) is tight for specific \textit{pure} states. But what happens if mixed states $\hat{\rho} = \sum_{x}\lambda_x|\lambda_x\rangle\langle\lambda_x|$ are considered? The mixed state introduces futher uncertainty which is not described by the entropies in the relation, thus making the lower bound loose. After a first intuition by \citet{rumin2011}, \citet{frank2012} understood that the intrinsic uncertainty of the mixed state could be included by using its von Neumann entropy and proved that the amended EUR
\begin{equation} \label{franck}
H(\hat{A})_{\rho} + H(\hat{B})_{\rho} \geq -\log_2c + S(\rho) 
\end{equation}
is tight. Their proof is taylored for continuous conjugated pairs connected by Fourier transform, but it has been later devised for discrete observables, too, by \citet{coles2017}. From the observation in Eq.~(\ref{purevsmixed}), we note that, if $\rho$ is maximally mixed, the new bound in Eq.~(\ref{franck}) is enhanced by $\log_2d$ with respect to the one in the Maassen-Uffink EUR, while, as expected, it is unchanged if $\rho$ is pure. It is worth remarking that, by adding the von Neumann entropy of $\rho$ and thus tightening the EUR, we have lost the state independence of the lower bound.

\subsection{Extension to POVMs}
\label{subsec34}
As we mentioned in section~\ref{subsubsec314}, the Born probabilities appearing in the EURs can be obtained by projecting on the eigenstates of the observables or, more in general, by means of a POVM, namely a set of positive operators $\Pi_{A}^{(a)}$ that sum to the identity, $\sum_{a \in \mathcal{A}}\Pi_A^{(a)} = \mathbb{I}$. However, note that the EURs inspected so far have been established in terms of projective measurements, thus implying the expression of the incompatibility factor in Eq.~(\ref{incompa}) as a function of the scalar product of the observable eigenstates. Here we remove the assumption of projective measurements and consider POVMs. This issue has first been addressed by \citet{krishna2002}. They found that the Maassen-Uffink EUR, with the POVMs $\Pi_A$ and $\Pi_B$ associated to the systems $A$ and $B$, holds with incompatibility $\tilde{c}$ given by
\begin{equation} \label{kri}
\tilde{c} \equiv \max_{a\in\mathcal{A},b\in\mathcal{B}}\left \lVert\sqrt{\Pi^{(a)}_A}\sqrt{\Pi^{(b)}_B}\right \rVert^2
\end{equation}
where $\lVert\cdot\rVert$ is the operator norm, identifying the largest singular value. This result has been obtained by converting the POVMs into projective measurements through Naimark extensions and then applying the Maassen-Uffink EUR. Interestingly, \citet{krishna2002} noted that, by setting $A=B$, we get a non trivial UR even for a single system, reading
\begin{equation}
H(\hat{A}) \geq -\log_2 \max_{a,a'\in\mathcal{A}} \left \lVert\sqrt{\Pi^{(a)}_A}\sqrt{\Pi^{(a')}_A}\right \rVert.
\end{equation}
This is due to the uncertainty intrinsic in the construction of a POVM. \\ \\
The result in Eq.~(\ref{kri}) has been generalized by \citet{rastegin2008} to Rényi entropies. \citet{tomamichel2015}, following a different path for the generalization to the POVMs, found a lower bound given by the incompatibility
\begin{equation}
\tilde{c}_T \equiv \min\left[\max_{b\in\mathcal{B}}\left\lVert\sum_{a\in\mathcal{A}}\Pi_{A}^{(a)}\Pi_{B}^{(b)}\Pi_{A}^{(a)}\right\rVert, \max_{a\in\mathcal{A}}\left\lVert\sum_{b\in\mathcal{B}}\Pi_{B}^{(b)}\Pi_{A}^{(a)}\Pi_{B}^{(b)}\right\rVert\right].
\end{equation}
This bound has been proved to be stronger than the one obtained with the incompatibility $\tilde{c}$ in Eq.~(\ref{kri}) by \citet{coles2014}.

\section{EURs with side information}
\label{sec5}
Finally, we turn to quantifiers that condition the uncertainty to the knowledge of one or more observers and inspect if uncertainty relations can still be established in this context. This knowledge of the observers is usually referred to as \textit{side information} or \textit{memory}, and it can be described either by a classical random variable or by a quantum state. 
\\
Therefore, in this framework we always have at least two systems: the observer system, $O$, where the observables of interest are measured, and the memory, $M$, where the a priori information is stored. We qualify the memory as \textit{classical} or \textit{quantum} depending on the correlations between $M$ and $O$. In addition, we need to consider the \textit{classical registers}, identified by auxiliary systems, correlated with the memory, where the information on the measurement outcomes is encoded. The joint system given by the register of the observable $\hat{X}$ and the memory $M$ is given by
\begin{equation} \label{cq}
R_{XM} = \sum_{x\in \mathcal{X}}p(x)|x\rangle\langle x| \otimes \rho_{M|x}
\end{equation}
which is known as \textit{classical-quantum state}, with $\rho_{M|a}$ the state of the memory conditioned on the outcome $a$. By using this description, we are compacting in the expression of the states the uncertainty on the measurement outcomes and the uncertainty on the state itself, namely we are expressing classical and quantum information with the von Neumann entropy. Therefore, hereon we will use the same notation $H(\cdot)_{\rho}$ to address both Shannon and von Neumann entropies. The same will be extended to their generalization to Rényi and smooth entropies.
\subsection{Classical memory}
\label{subsec51}
In the case of classical memory, the correlations between memory and observer must be classical, as the one between memory and registers described by the state in Eq.~(\ref{cq}). Here we show that the EURs for Shannon unconditional entropies presented in sections~\ref{subsec31},~\ref{subsec32} and~\ref{subsec33} hold with the same bound also for the corresponding quantities conditioned on a memory system and yield stronger relations. \\
Let us consider the EUR in Eq.~(\ref{geneur}) for the set of complementary observables $\{\hat{A}_j\}_{j}$, with the projective measurements $\Pi_{A_j}=\{|a\rangle\langle a|\}_{a\in\mathcal{A}}$, on the observer system $O$. We want to devise a corresponding inequality for $H(\hat{A}_j|M)_{\rho_{A_jM}}$, with
\begin{equation} \label{cqclassmem}
\rho_{A_jM} = \sum_{m\in\mathcal{M}}p_M(m)\rho_{A_j|m}\otimes |m\rangle\langle m|_{M}
\end{equation}
and the memory $M$ a classical random variable taking values from a set $\mathcal{M}$ with distribution $p_{M}(m)$.
The probabilities $p_M(m)$ are defined after the joint distribution
\begin{equation}
p_{A_jM}(a,m) = p_M(m) \, p_{A_j}(a|m) = p_M(m)\,\text{Tr}[\rho_{A|m}\Pi_{A_j}^{(a)}].
\end{equation}
The expression of the classical-quantum state $\rho_{A_jM}$ in Eq.~(\ref{cqclassmem}) is determined by the hypothesis of classical memory, i.e. it is not entangled. It is simple to show that the conditional entropies can be expressed as
\begin{equation}
H(\hat{A}_j|M)_{\rho_{A_jM}} = \sum_{m}p_M(m)H(\hat{A}_j)_{\rho_{A_j|m}},
\end{equation}
which follows from Theorem~\ref{JET} in section~\ref{subsubsec314} and from the definition of conditional entropy. Note that $H(\hat{A}_j)_{\rho_{A_j|m}}$ is often indicated as $H(\hat{A}_j|M=m)$. If we take the sum over all the observables $\hat{A}_j$, we get
\begin{equation}
\sum_j H(\hat{A}_j|M)_{\rho_{A_jM}} = \sum_{m}p_{M}(m)\sum_jH(\hat{A}_j)_{\rho_{A_j|m}}.
\end{equation}
However, the EUR in Eq.~(\ref{geneur}) holds for every state $\rho_{A_j|m}$ with $m$ identifying the classical memory $M$. But then
\begin{equation}
\sum_jH(\hat{A}_j) \geq \mathcal{B} \Longrightarrow \sum_jH(\hat{A}_j|M)_{\rho_{A_jM}}\geq\mathcal{B}\sum_{m}p_{M}(m) = \mathcal{B}.
\end{equation}
Each of the EURs for Shannon entropies imply a corresponding uncertainty relation for the related conditional quantities. 
Note that the relation for conditional entropies is generally tighter since conditioning never increases the entropy, i.e. $H(\hat{A}_j|M) \leq H(\hat{A}_j)$, with equality only if $O$ and $M$ are uncorrelated.

\subsection{Quantum memory}
\label{subsec52}
Now we turn to the case where the observer $O$ and the memory $M$ can be entangled. Does the complementarity of the observables imply an uncertainty relation in the presence of quantum correlations, too? Before delving into formal results, let us consider the case where the systems $O$ and $M$ are described by a maximally entangled state, such as the Bell state
\begin{equation} \label{qom}
|\Phi\rangle_{OM} = \frac{1}{\sqrt{2}}\left(|0\rangle_O\otimes|0\rangle_M + |1\rangle_O\otimes|1\rangle_M \right).
\end{equation}
Then, every outcome found with a measurement on $O$ would determine a perfectly correlated result on $M$ on the basis $\{|0\rangle, |1\rangle\}$. However, this is not particularly interesting in terms of uncertainty relations, since no complementary measurements are involved yet. Moreover, note that the same correlations may be obtained with a classical memory, by using the state
\begin{equation} \label{clom}
\rho_{OM} = \frac{1}{2}\left(|0\rangle\langle 0|_O \otimes \rho_{M|0} +|1\rangle\langle 1|_O \otimes \rho_{M|1}  \right),
\end{equation} 
and, again, measuring on $\{|0\rangle, |1\rangle\}$. The phenomenology changes drastically if one performs measurements also in the complementary basis $\{|+\rangle,|-\rangle\}$, with $|+\rangle = (|0\rangle + |1\rangle)/\sqrt{2}$ and $|-\rangle = (|0\rangle - |1\rangle)/\sqrt{2}$. While in the classical case in Eq.~(\ref{clom}) the memory loses all the information about the outcome obtained by the observer, the quantum memory keeps the same correlation with $O$, since Eq.~(\ref{qom}) can be equivalently rewritten in terms of the basis $\{|+\rangle,|-\rangle\}$, namely
\begin{equation} \label{qom}
|\Phi\rangle_{OM} = \frac{1}{\sqrt{2}}\left(|+\rangle_O\otimes|+\rangle_M + |-\rangle_O\otimes|-\rangle_M \right).
\end{equation}
Therefore, we expect that, if the memory is maximally entangled with the observer, the uncertainty relations yield trivial bounds.
\\
EURs in the presence of a quantum memory has been firstly addressed for quantum channels by \citet{christandl2005}, and for tripartite states (see section~\ref{subsec53}) by \citet{renes2009}. \citet{berta2010} has explored the case of bipartite systems and proved the following result.
\begin{theorem}[EUR for bipartite states with quantum memory] \label{berth}
Consider two observables $\hat{A}$ and $\hat{B}$, with eigenstates $\{|a\rangle\langle a|\}_{a\in \mathcal{A}}$ and $\{|b\rangle\langle b|\}_{b\in \mathcal{B}}$ and outcomes in the sets $\mathcal{A}$ and $\mathcal{B}$, respectively. The observables are projectively measured in the system $O$. Be $\rho_{AM}$ the classical-quantum state
\begin{equation}
\rho_{AM} = \sum_{a}|a\rangle\langle a|_A \otimes {\rm Tr}_O[(|a\rangle\langle a|_O\otimes \mathbb{I}_M)\rho_{OM}]
\end{equation}
where $A$ is the auxiliary register for the observable $\hat{A}$, and similarly for $\rho_{BM}$ with the auxiliary register $B$. Then, for every state $\rho_{OM}$ and every pair of observables $\hat{A}$ and $\hat{B}$,
\begin{equation} \label{berteur}
H(\hat{A}|M) + H(\hat{B}|M) \geq -\log_2c +H(O|M),
\end{equation}
where $c$ is the incompatibility factor defined in Eq.~(\ref{incompa}).
\end{theorem}
Here, the uncertainty quantifiers are $H(\hat{A}|M)$ and $H(\hat{B}|M)$, describing the ignorance about the measurement outcomes of $\hat{A}$ and $\hat{B}$ given the information in the memory system $M$. The conditonal entropy $H(O|M)$, which is part of the bound in the EUR in Eq.~(\ref{berteur}), is a measure of the correlations between the systems $O$ and $M$ and, as mentioned in section~\ref{subsubsec314}, conditional entropies, in the presence of quantum correlations, are negative. In particular, if $d$ is the dimension of the system $O$, we have $H(O|M) \geq -\log_2 d$, tight for maximally entangled states, while, in the case of separable states, $H(O|M) \geq 0$. Then, we observe that
\begin{itemize}
\item the lower bound in the EUR of Eq.~(\ref{berteur}) is \textit{state-dependent} through $H(O|M)$,
\item as expected, if $\rho_{OM}$ is maximally entangled and the bases of $\hat{A}$ and $\hat{B}$ are mutually unbiased, the EUR reduces to the trivial inequality $H(\hat{A}|M) + H(\hat{B}|M) \geq 0$ (and, if the bases are not mutually unbiased, the bound is even negative),
\item in general, quantum correlations improve the ultimate limit to precision that can be achieved when dealing with complementary observables,
\item if the memory system $M$ is discarded, the EUR simplifies to the state-dependent Maassen-Uffink EUR in Eq.~(\ref{franck}), $H(\hat{A}) + H(\hat{B}) \geq -\log_2c + H(\rho_{O})$.
\end{itemize}
From these considerations, we see that the EUR in Eq.~(\ref{berteur}) can also be read as an entanglement witness \citep{berta2010}: the inequality $\mathcal{B} < -\log_2c$ is a necessary and sufficient condition for the state $\rho_{OM}$ presenting quantum correlations. This application has been investigated in lab by \citet{li2011} and \citet{prevedel2011}, while providing experimental verification of Eq.~(\ref{berteur}).
\\
The proof of Theorem~\ref{berth} provided by \citet{berta2010} is entirely based on manipulations of the smooth min- and max-entropy. Later, simpler proofs have been proposed by \citet{coles2011,coles2012} and \citet{frank2013}. 

\subsection{Tripartite systems}
\label{subsec53}
When in communication theory we want to keep into account the security and the privacy of the transmitted information against an adversary, we need to consider at least three parties: sender, receiver and eavesdropper. This issue can be addressed by extending the case explored in section~\ref{subsec52} to the presence of two quantum memories, $M_{1}$ and $M_2$. We have still an observer $O$ measuring either the observable $\hat{A}$ or $\hat{B}$, but now the register of $\hat{A}$ is correlated with $M_1$ and the one of $\hat{B}$ to $M_2$. Given this setting, we inquire if the complementarity of $\hat{A}$ and $\hat{B}$ sets a bound on the sum of the uncertainties related to the $\hat{A}$ and $\hat{B}$ outcomes given the information stored in $M_1$ and $M_2$, i.e. $H(\hat{A}|M_1)$ and $H(\hat{B}|M_2)$, respectively. \citet{renes2009} have addressed this point and conjectured a generalization of the Maassen-Uffink EUR, which we present in the following theorem.
\begin{theorem}[EUR for tripartite states with two quantum memories] \label{rennth}
Consider two observables $\hat{A}$ and $\hat{B}$, with outcomes in the sets $\mathcal{A}$ and $\mathcal{B}$, respectively, and two POVMs $\{\Pi_A^{(a)}\}_{a\in\mathcal{A}}$ and $\{\Pi_B^{(b)}\}_{b\in \mathcal{B}}$. Consider a system $O$ where the POVMs are performed and two side systems $M_1$ and $M_2$. Then, for every tripartite state $\rho_{OM_1M_2}$,
\begin{equation} \label{renes}
H(\hat{A}|M_1) + H(\hat{B}|M_2) \geq -\log_2\tilde{c}.
\end{equation}
where $\tilde{c}$ is the incompatibility factor defined in Eq.~(\ref{kri}).
\end{theorem}
\citet{renes2009} proved it in the specific case where $\hat{A}$ and $\hat{B}$ are related by Fourier transform, which implies that they are mutually unbiased and $c = 1/d$. In particular,
\begin{equation} 
\hat{A} = \hat{F}\hat{B}\hat{F}^{\dagger} \quad {\rm with} \quad \hat{F} = \frac{1}{\sqrt{d}}\sum_{n,m}e^{-2\pi i \frac{nm}{d}}|n\rangle \langle m|.
\end{equation}
Eventually, Eq.~(\ref{renes}) has been demonstrated for generic observables by \citet{berta2010}. They showed that Eq.~(\ref{renes}) can be understood as a reformulation for tripartite systems of their result in Eq.~(\ref{berteur}) for bipartite states. The derivation just requires to notice that for a pure state $\rho_{OM_1M_2}$ one has $H(\hat{A}, M_1) = H(\hat{A},M_2)$ and $H(O,M_1) = H(M_2)$.
\\
While in the bipartite case the EUR is reduced to a trivial inequality if the systems $O$ and $M$ are maximally entangled, a proper uncertainty relation is always possible for tripartite states. This follows from a basic property of entangled states, known as \textit{monogamy of entanglement} \citep{coffman2000}: if two systems are maximally entangled, they cannot be entangled with a third party, and, in general, the entanglement of $O$ and $M_1$ affects the entanglement of $O$ and $M_2$. The monogamy of entanglement is strictly connected with the \textit{no-cloning theorem} \citep{wootters1982}, which states that it is in general not possible to copy quantum information. Therefore, the EUR in Eq.~(\ref{berteur}) can also be read as a consequence of the no-cloning: the memory systems $M_1$ and $M_2$ cannot hold copies of each other quantum information. On the other hand, the no-cloning is a direct consequence of the existence of complementary observables \citep{nielsen2010}. Hence, we see that the complementarity plays at least two different roles in assessinn the uncertainty in the EURs: through the superposition of the bases of the observables, it defines the incompatibility factor bounding the sum of the entropies, and, additionally, for systems with more than two parties, it prevents from storing in one memory all the information obtained from measurements and other memories.
\\
\citet{muller2013} has extended Theorem~\ref{rennth} to Rényi entropies, with the notion of conditional Rényi entropy introduced in Definition~\ref{sqre}. The generalized EUR reads
\begin{equation}
H_{\alpha}(\hat{A}|M_1) + H_{\beta}(\hat{B}|M_2) \geq -\log_2c,
\end{equation}
with $\alpha^{-1}+\beta^{-1} = 2$.
Interestingly, the same result has been obtained by \citet{coles2012}, but using a different definition of conditional Rényi entropy, following \citet{tomamichel2009}.
\\
Finally, \citet{tomamichel2011} have generalized the EUR in Eq.~(\ref{renes}) to smooth min- and max-entropy, i.e.
\begin{equation} \label{eurtomren}
H_{\text{min}}^{\varepsilon}(\hat{A}|M_1) + H_{\text{max}}^{\varepsilon}(\hat{B}|M_2) \geq -\log_2\tilde{c},
\end{equation}
with, as in Theorem~\ref{rennth}, the incompatibility factor $\tilde{c}$ given by Eq.~(\ref{kri}). We will see in section~\ref{subsec61} that this EUR defines the security of a communication scheme with discrete variables and finite resources.
\section{Applications}
\label{sec6}
\subsection{Cryptography}
\label{subsec61}
The existence of complementary observables, together with quantum correlations, determines the physical foundations of security in quantum cryptography. 
\\
There, two parties, Alice and Bob, want to share a private communication, secure against a possible adversary, the eavesdropper. To this aim, Alice and Bob must compute a secret key that the sender, Alice, will use to encrypt the message and the receiver, Bob, to decrypt it. If the key is actually secret, i.e. known to Alice and Bob only, and the key satisfies the requirements of the Vernam cipher (an infinite string of random numbers), then the communication is secure, as the Vernam cipher is perfect \citep{shannon1945,shannon1948}. However, in classical cryptography we have that the process of key generation is probelmatic: it is still a communication task but it cannot be actually secure by definition since, at that level, there is no key to encrypt the message yet. A solution to this issue is offered by quantum mechanics: the intrinsic uncertainty due to the complementary properties that we have characterized so far can be used to ground the notion of security on a physical base. The quantum uncertainty prevents from devising a universal unitary copying operation. Moreover, it implies a level of randomness that cannot be achieved by classical algorithms, thus better realizing a Vernam cipher.
\subsubsection{Quantum key distribution}
The area of cryptography addressing the security of the key generation and sharing is known as \textit{quantum key distribution} (QKD). The first and most famous QKD protocols are the BB84, by \citet{bennett1984}, the E91, by \citet{ekert1991}, and the BBM92, by \citet{bennett1992}. Here we will mainly focus on the BB84 and BBM92 protocols, together with their extensions to high dimensions, since these are the ones critically based on quantum uncertainty, while the E91 exploits the violation of Bell-like inequalities.
\\
\\
The BB84 protocol is the very beginning of QKD. This is a \textit{prepare and measure} scheme, where Alice prepares a quantum state, sends it through an insecure quantum channel and Bob measures it. The crucial point in this scheme is that Alice can choose between two complementary bases for the preparation of the state, say the eigenstates of the Pauli matrices $\hat{Z}$ and $\hat{X}$, i.e. $\{|0\rangle, |1\rangle\}$ and $\{|+\rangle,|-\rangle\}$, with $|\pm\rangle \equiv (|0\rangle \pm |1\rangle)/\sqrt{2}$. Bob, as well, can choose to perform a projective measurement in one of the two bases. The bits are transmitted through the insecure quantum channel. This procedure is repeated for a pre-selected number of times. At every round, Alice and Bob write down the basis that they chose and the state that they prepared/measured. 
\\
At the end of the distribution, Alice and Bob, by means of local operations and classical communication, publicly announce their choices of bases and discard the rounds where the choice is different. This part of the protocol is named \textit{sifting}. In the absence of noise and of an eavesdropper, the bits corresponding to the states they prepared and measured, round by round, should be perfectly correlated, and they could use them as a key. However, a fraction of the bits can be spolit by the noise introduced in by the quantum channel and the errors in preparation and measurement. Crucially, in this scheme \textit{a potential eavesdropper cannot avoid to add further noise}, since all the information that she can obtain implies a disturbance of the state, that Alice and Bob can quantify. This is done in the \textit{parameter estimation} step, where they share publicly a fraction of their bits and compute the \textit{quantum bit error rates} (QBERs). If these are larger than a pre-selected threshold, established by estimating the global noise affecting the apparatus and/or the minimal number of bits that the eavesdropper needs to reconstruct the whole key, they abort the protocol. If not, they apply an error correcting code to get the same bit string (\textit{information reconciliation}), which they eventually map into a safer key (\textit{privacy amplification}).
\\
\\
The BBM92 protocol is the \textit{entanglement-based} version of the BB84. In this scheme, there is an external untrusted source of entangled states. One party is sent to Alice and one to Bob and they \textit{both} measure one of two complementary observables. If the state is maximally entangled and bipartite, then, in the absence of noise and adversaries, after sifting they should have perfectly correlated bit strings. In this scenario, we test the security of the protocol by giving to the eavesdropper the maximal power, i.e. the control of the untrusted source. The best she can do is to distribute a tripartite entangled state and keep one of the parties. Moreover, she is assumed to hold a quantum memory. However, due to the monogamy of entanglement, this operation will result in additional noise in Alice's and Bob's outcomes. Therefore, hereon they apply the same procedure described above: parameter estimation, information reconciliation and privacy amplification.
\\
The E91 protocol is entanglement-based too, but it exploits a different method to assess the security. We remark that the context and the quantities involved in the BBM92 should sound familiar after reading section~\ref{subsec53}: we will see in the following of the EURs in the security proof of this protocol.
\\
\\
But first, one may ask what is the relation between BB84 and BBM92, except for some common steps in the procedure. From the experimental point of view, the implementation of the latter is simpler, since it does not require the generation of entangled states. Conversely, the formalization of the security proof is straightforward for the BBM92, and it is more involved for the BB84. The formal relation between a prepare-and-measure protocol and the corresponding entanglement-based version has been assessed by \citet{tomamichel2017}: \textit{the security of the prepare-and-measure protocol follows from the security of the entanglement-based version, given that some assumptions on preparation and measurement are fulfilled}. Therefore, in many cases one can use the security proof of the entanglement-based protocol to infer the security of the prepare-and-measure one. If the assumptions for the equivalence are not satisfied, then using the entanglement-based security proofs for the prepare-and-measure scenario gets the protocol exposed to loopholes that can be exploited by the eavesdropper. 

\subsubsection{The EUR for smooth entropies in QKD}
We denote $R_A$ and $R_B$ the random variables describing the \textit{raw keys} of Alice and Bob, namely the partially correlated keys that they have before information reconciliation. Let $E$ be the random variable for the information got by the eavesdropper. To avoid confusion, we refer to the complementary observables as $\hat{X}$ and $\hat{Z}$ here, with $Q_X$ and $Q_Z$ the pertaining QBERs.
\\
One of the most relevant quantities to be optimized to assess the security of a QKD protocol is the \textit{secret key rate} $r$, i.e. the number of secret bits $n$ over the total number of rounds $N$. In the limit of an infinite number of rounds, \citet{devetak2005} found that it asymptotically converges to
\begin{equation} \label{akr}
r_{\infty} \equiv H(R_A|E) - H(R_A|R_B)
\end{equation}
which has an immediate interpretation: it is the difference between the uncertainty of Eve about Alice's raw key and the uncertainty of Bob. It is called \textit{asymptotic key rate}. In the case of the BBM92 protocol, by computing the conditional entropies in Eq.~(\ref{akr}), one finds \citep{grasselli2021}
\begin{equation}
r_{\infty}^{{\rm (BBM92)}} = 1-h(Q_X) - h(Q_Z),
\end{equation}
where $h(\cdot)$ is the binary Shannon entropy defined in Eq.~(\ref{binaryentropy}). Note that this is a simple expression depending on the QBERs only, i.e. the quantities that are experimentally observed in the protocol.
\\
In the unrealistic asymptotic scenario described by $r_{\infty}$ we miss the security enhancement provided by the information reconciliation and privacy amplification steps. We can take into account their contributions if we consider a finite number of rounds. The information reconciliation fixes the \textit{correctness} of the key, i.e. Alice and Bob share the same key at the end of this step. To this aim, one of the two parties, say Alice, sends the minimal amount of information $n^{(\varepsilon)}(A|B)$ such that Bob can correctly guess her key up to a failure probability $\varepsilon$. But, as we have mentioned in section~\ref{subsubsec314}, this is exactly the definition of smooth max-entropy. Therefore, as noted by \citet{konig2009},
\begin{equation}
n^{\varepsilon}(A|B) = H_{\rm max}^{\varepsilon}(A|B) + o(\log_2(1/\varepsilon)).
\end{equation}
The privacy amplification fixes the \textit{secrecy} of the key shared by Alice and Bob.
\begin{definition}[Secrecy]
A key $K$ is defined to be secret with respect to an adversarial party $E$ if it is uniformly distributed and independent of $E$.
\end{definition}
The secrecy of $K$ is achieved in privacy amplification by applying a random function $f$ to $K$, obtaining $L = f(K)$ such that
\begin{equation} \label{condtr}
D_T\left(\rho_{LE}, \frac{1}{|\mathcal{L}|}\sum_{l\in \mathcal{L}}|l\rangle \langle l|\otimes\rho_E \right) \leq \tilde{\varepsilon},
\end{equation}
where $\mathcal{L}$ is the set of all possible final keys $l$ (including $L$) shared by Alice and Bob, $\tilde{\varepsilon}$ is a positive parameter smaller than one and 
\begin{equation}
D_T(\rho,\sigma) \equiv \frac{1}{2}{\rm Tr}\left[\sqrt{(\rho - \sigma)^2}\right]
\end{equation}
is the trace distance. Therefore, the final secret key is the number of bits extracted from $K$ that are $\varepsilon$-indistinguishable in the trace distance from the formal definition of secret bits, up to an error probability. By using a Lemma, known as \textit{quantum leftover hash lemma} \citep{tomamichel2011b}, that connects the trace distance in Eq.~(\ref{condtr}) with the smooth min-entropy and the secret key, it is possible to show the following.
\begin{lemma}[Secret key rate] \label{rsmooth}
Consider the classical-quantum state
\begin{equation}
\rho_{KE} = \sum_{k\in\mathcal{K}}p_k|k\rangle\langle k|_K \otimes \rho_{E|k}
\end{equation}
with $\{|k\rangle\}_{k\in\mathcal{K}}$ a set of orthonormal states representing Alice's possible keys with probability distribution $\{p_k\}_{k\in \mathcal{K}}$. It describes an insecure key string in the register $K$ and the side information of the eavesdropper $E$. For a finite number $N$ of rounds, the key rate, with key $\tilde{\varepsilon}$-indistinguishable from a secret key, that can be extracted from $K$, up to an error probability $\varepsilon$, is quantified by the smooth min-entropy $H_{\rm min}^{\varepsilon}(K|E)$, with $\varepsilon < \tilde{\varepsilon}$, as
\begin{equation} 
r \leq \frac{1}{N}\left[H_{\rm min}^{\varepsilon}(K|E) + 2\left(1+\log_2(\tilde{\varepsilon} - \varepsilon)\right)\right].
\end{equation}
\end{lemma}
The security parameters, such as $\varepsilon$ and $\tilde{\varepsilon}$, are eventually used to define the \textit{composable security} of the protocol, i.e. they are optimized and their sum is upper bounded by a term $\varepsilon_{\rm tot}$ which defines the protocol as $\varepsilon_{\rm tot}$-secure \citep{renner2008,tomamichel2012}.
\\
\\
We infer from Lemma~\ref{rsmooth} that the main task of a discrete-variable QKD protocol is to provide the tighest estimation of $H_{\rm min}^{\varepsilon}(K|E)$. In particular, it is important to lower bound this quantity, in order to assess the minimum number of $\varepsilon_{\rm tot}$-secure bits of the key. We have seen in section~\ref{subsec53} that there is an EUR that can solve this task, i.e. Eq.~(\ref{eurtomren}). Without loss of generality, we can assume that the observable $\hat{Z}$ is used for the generation of the key and $\hat{X}$ to check the QBERs in the parameter estimation step. Then, we can rewrite the EUR by \citet{tomamichel2011} as follows
\begin{equation} \label{eurqkd}
H^{\varepsilon}_{\rm min}(\hat{Z}|E) + H^{\varepsilon}_{\rm max}(\hat{X}|B) \geq -\log_2\tilde{c}.
\end{equation}
The origin of this bound stems from the very notion of complementarity and is, thus, the tighest. Next, one needs to lower bound the smooth max-entropy to achieve the minimal smooth min-entropy. Given the QBERs, this problem can be reduced to bounding the entropy of a classical probability distribution, which can be done by exploiting classical results for sampling without replacement. \citep{tomamichel2012} bounded the smooth max-entropy by a function of the estimated QBER in the check basis, $\hat{X}$. This provides a lower bound on the smooth min-entropy which ultimately allows to estimate the secret key rate in terms of the security parameters. 
\\
The drawback of using the EUR in Eq.~(\ref{eurqkd}) is the number of complementary observables that can be considered, which is restricted to two. So far, no extensions to many observables are known, except for an attempt by \citet{wang2021}. This is an interesting direction since it has been proved that using more than two mutually unbiased bases helps to improve the maximum tolerable error rate \citep{bruss1998,bradler2016,wyderka2025}.
\subsection{Metrology}
\label{subsec62}
One of the most direct implications of the URs is the ultimate limit on the precision of a joint measurement of complementary observables. However, there are physical quantities that are not described by operators in quantum mechanics, such as time and phases. A famous example is the well-known time-energy UR by Heisenberg, where the energy is associated to the Hamiltonian operator, but time is described as a paramerer. As such it has no quantum uncertainty. Indeed the time-energy uncertainty relation must be interpreted as a quantum speed limit: the minimum \textit{time interval} $\Delta t$ it takes for a system with energy uncertainty $\Delta E$ is lower bounded by $\Delta t \geq \hbar/(2\Delta E)$ \citep{aharonov1961,giovannetti2003}. But then, how can one speak of complementary properties if the involved quantities are not given as observables? These issues led to investigate how these quantities can be \textit{estimated} \citep{helstrom1976,holevo1982}. 
\\
In order to address the problem of estimating a parameter $\phi$, we need to find an estimator $\bar{\phi} = \bar{\phi}\,(x_1,x_2,..., x_N)$, which is a function that maps the $N$ outcomes of a set of measurement into the parameter space. The precision of the estimation provided by the $N$ measurements is typically quantified either by the variance of the estimator \citep{paris2009}
\begin{equation}
\text{var}(\bar{\phi}) \equiv \langle (\bar{\phi} - \phi)^2\rangle
\end{equation} 
or by the second moment of the statistical distance \citep{braunstein1994}
\begin{equation}
\delta \bar{\phi} \equiv \frac{\bar{\phi}}{\left| d\langle\bar{\phi}\rangle/d\phi \right|} - \phi.
\end{equation}
In classical estimation theory, the most precise estimation that can be achieved in terms of both, $\text{var}(\bar{\phi})$ or $\langle (\delta \bar{\phi})^2\rangle$, is established by the \textit{Cramér-Rao bound} \citep{cramer1946}
\begin{equation} \label{crbound}
\langle (\delta \bar{\phi})^2 \rangle \geq \frac{1}{N\,F(\phi)},
\end{equation}
where $F(\phi)$ is the \textit{Fisher information}, which is the variance of the gradient of the log-likelihood function with respect to the parameter (namely, the \textit{score}), i.e.
\begin{equation} \label{fisher}
F(\phi) \equiv \int dx\,p(x|\phi)\left(\frac{\partial\ln p(x|\phi)}{\partial\phi}\right)^2.
\end{equation}
$F(\phi)$ measures the information that an estimator $\bar{\phi}\,(x_1,...,x_N)$ can provide about the parameter $\phi$ given the conditional distribution $p(x|\phi)$. Optimal estimators are the ones that saturates the Cramér-Rao bound.
\\
In quantum theory, the probabilities $p(x|\phi)$ are identified by the Born rule as 
\begin{equation}
p(x|\phi) = \text{Tr}[\rho_{\phi}\Pi_x],
\end{equation}
where $\Pi_x$ is the element of a POVM yielding $x$ as an outcome and $\rho_{\phi}$ is the state where the parameter $\phi$ has been encoded. In order to devise a quantum Cramér-Rao bound, \citet{helstrom1976}, \citet{holevo1982} and \citet{braunstein1994} have introduced the \textit{symmetric logarithmic derivative}.
\begin{definition}[Symmetric Logarithmic Derivative]
We define the symmetric logarithmic derivative as the self-adjoint operator $\hat{L}_{\phi}$ such that
\begin{equation}
\frac{1}{2}\left\{\hat{L}_{\phi},\rho_{\phi}\right\} = \frac{\partial\rho_{\phi}}{\partial \phi}.
\end{equation}
\end{definition}
This operator allows us to compactly express the derivative of $p(x|\phi)$ with respect to the parameter $\phi$ as
\begin{equation}
\frac{\partial p(x|\phi)}{\partial{\phi}} = \text{Tr}\left[\frac{\partial \rho_{\phi}}{\partial{\phi}}\Pi_x\right] = \text{Re}\left(\rho_{\phi}\Pi_x\hat{L}_{\phi}\right),
\end{equation}
and, hence, to rewrite the Fisher information in Eq.~(\ref{fisher}) as
\begin{equation} \label{fisher2}
F(\phi) = \int dx\,\frac{\text{Re}\left(\text{Tr}[\rho_{\phi}\Pi_x\hat{L}_{\phi}]\right)}{\text{Tr}[\rho_{\phi}\Pi_x\hat{L}_{\phi}]}.
\end{equation}
The quantum formalism unveals that the Fisher information can be further optimized over the choice of the POVM $\Pi =\{\Pi_x\}_x$. This optimization \citep{yuen1973,helstrom1974,braunstein1994,braunstein1996} eventually leads to an inequality tighter than Eq.~(\ref{crbound}).
\begin{theorem} [Quantum Cramér-Rao Bound]
Given a parameter $\phi$ and an estimator $\bar{\phi}(x_1,x_2,...,x_N)$, function of $N$ measurement outcomes, the ultimate bound on the second moment of the statistical distance $\langle (\delta \bar{\phi})^2 \rangle$ is given by
\begin{equation}
\langle (\delta \bar{\phi})^2 \rangle \geq \frac{1}{N\,G(\phi)}
\end{equation}
where
\begin{equation}
G(\phi) \equiv {\rm Tr}[\rho_{\phi}\hat{L}_{\phi}^2]
\end{equation}
is defined as Quantum Fisher Information and upper bounds the Fisher information in Eq.~(\ref{fisher2}), i.e. $G(\phi) \geq F(\phi)$.
\end{theorem}
A quantum measurement provides an optimal estimation of the parameter $\phi$ if it is described by a POVM whose Fisher information equals $G(\phi)$. Importantly, the quantum Fisher information depends on the quantum statistical model given by $\rho_{\phi}$ only.
\\
As the quantum Cramér-Rao bound establishes the ultimate limit to the precision of parameter estimation, one can see if this has any connection with some UR. \citet{braunstein1996} have found that this is the case.
\begin{theorem}[Parameter-based UR]
Consider the state $\rho = \sum_{n}p_n|n\rangle\langle n|$, with $\sum_np_n = 1$, and the Hamiltonian $\hat{H}$ generating the evolution described by
\begin{equation}
\frac{\partial \rho}{\partial \phi} = \sum_n\frac{dp_n}{d\phi}|n\rangle\langle n| - i[\hat{H}, \rho]
\end{equation}
and, for infinitesimal variations,
\begin{equation}
\rho_{\phi + d\phi} = \sum_n(p_n + dp_n)e^{-i\hat{H}d\phi}|n\rangle\langle n| e^{i\hat{H}d\phi}.
\end{equation}
Assume that the eigenvalues of the state $\rho$ do not depend on the parameter, i.e. $dp_n/d\phi=0$, implying $\partial\rho/\partial\phi = -i[\hat{H},\rho]$. Then,
\begin{equation} \label{urbrau}
\langle (\delta\phi)^2 \rangle \langle (\delta\hat{H})^2 \rangle \geq \frac{1}{4N}.
\end{equation}
\end{theorem}
This result holds also if the statistical distance of the Hamiltonian is replaced by its variance, but it is less tight. The UR in Eq.~(\ref{urbrau}) stems directly from the quantum Cramér-Rao bound through the identity $G(\phi) = 4\langle (\delta\hat{H})^2 \rangle$. Therefore, it is not necessary for a pair of observables satisfying an UR to be identified as operators. On the other hand, the ultimate limit given by the quantum Cramér-Rao bound can be interpreted as a parameter-based UR.
\\
\\
This deep connection between quantum parameter estimation and URs has opened new perspectives. Crucially, it led \citet{giovannetti2006} to inspect the cases where quantum effects can enhance the precision of parameter estimation, a framework known as \textit{quantum metrology}. The scaling $1/N$ following from the UR in Eq.~(\ref{urbrau}) has been identified as \textit{Heisenberg scaling} and found to be an enhancement in precision by $\sqrt{N}$ with respect to the bound obtained through the central limit theorem, i.e. the \textit{standard quantum limit} \citep{giovannetti2004}.

\section{Conclusions and Outlooks}
We have addressed the notion of uncertainty in quantum mechanics. We have shown how it has been understood and described so far by following two approaches: the description in terms of variance and the one in terms of entropy. The variance-based approach is essential to introduce the first formulations of the URs and is nowadays exploited in many relevant applicationThe entrropy-based approach allows to overcome some issues of the variance-based URs and provide a rich description of quantum uncertainty, due to the deep structure of the Rényi entropies, the generalization to smooth entropies and the natural implementation of side information. Moreover, due to the practical origin of these notions of entropy to accomplish information-theoretic tasks, the EURs have immediately found application in many contexts, in particular quantum computation, randomness extraction and quantum cryptography.

\appendix
\section{MUBs in terms of Hadamard matrices}
\label{app1}
Let us consider a set of mutually unbiased bases in a generic dimension $d$ and represent each basis as a matrix where the columns are the vector of the basis. Due to the mutual unbiasedness, we obtain a set of complex Hadamard matrices where the entries have all modulus $1/\sqrt{d}$. Therefore, the classification of the mutually unbiased bases can be reduced to the classification of complex Hadamard matrices. \cite{brierly2010} have fulfilled this task for dimensions from two to five.
\\
The classification is much simplified by the existence of equivalence classes of mutually unbiased bases. In particular, two sets of mutually unbiased bases are equivalent if the same unitary transformation maps one bases into the other. This notion of equivalence allows to define a \textit{standard form}, where one of the bases is always the identity and the others are the Hadamard matrices whose vectors are mutually unbiased with the vectors of the identity.

\section{MUBs in terms of generalized Pauli matrices}
\label{app2}
The unitary operators of the Heisenberg-Weyl group are mutually unbiased, thus providing a classification for these bases. The Heisenberg-Weyl operators $\hat{X}^{\alpha}\hat{Z}^{\beta}$ generalize the Pauli matrices to generic dimensions. In particular, 
\begin{align}
&\hat{X} \equiv \sum_j |j\rangle\langle j-1| \\
&\hat{Z} \equiv \sum_j \omega^j |j\rangle \langle j|
\end{align}
with $\omega = \exp{(2\pi i/d)}$. While in the case of prime dimensions $d$ the eigenbases of the $d+1$ operators $\hat{Z}$ and $\hat{X}\hat{Z}^k$ with $k\in\{0,1,\ldots d-1\}$ are known to form a maximal set of $d+1$ MUBs \citep{wootters1988,englert2001,durt2010}, for all the other cases, just the bases of $\hat{Z}$, $\hat{X}$ and $\hat{X}\hat{Z}$ can be proved to be mutually unbiased.



\bibliographystyle{elsarticle-harv} 




%

\end{document}